\input epsf
\documentclass[nohyper,notoc]{article} 
\usepackage{axodraw}
\usepackage{epsfig}

\def\beq{\begin{equation}}
\def\eeq{\end{equation}}
\def\bea{\begin{eqnarray}}
\def\eea{\end{eqnarray}}
\def\bq{\begin{quote}}
\def\eq{\end{quote}}
\def \lsim{\mathrel{\vcenter
     {\hbox{$<$}\nointerlineskip\hbox{$\sim$}}}}
\def \gsim{\mathrel{\vcenter
     {\hbox{$>$}\nointerlineskip\hbox{$\sim$}}}}
\def\gappeq{\mathrel{\rlap {\raise.5ex\hbox{$>$}}
{\lower.5ex\hbox{$\sim$}}}}
\def\lappeq{\mathrel{\rlap{\raise.5ex\hbox{$<$}}
{\lower.5ex\hbox{$\sim$}}}}

\def\mnu{[m_{\nu}]_{IJ}}

\def\meg{\mu \to e \gamma}
\def\tmg{\tau \to \mu \gamma}
\def\m3e{\mu \to e \bar{e} e}
\def\mec{\mu N  \to e N'}
\def\a{\alpha}
\def\b{\beta}

\def\m{\mu}

\def\bU{\bar{\Upsilon}}
\def\bUI{\bar{\Upsilon}_I}
\def\tWe{\widetilde{W}_e}
\def\tWu{\widetilde{W}_u}
\def\tWd{\widetilde{W}_d}
\def\We{W_e}

\begin{document}

\renewcommand{\thefootnote}{\fnsymbol{footnote}}
\begin{flushright}
LPT-ORSAY/10-72
  \\
\end{flushright}
\vskip 5pt
\begin{center}
{\Large {\bf ``Minimal Flavour Violation'' for  Leptoquarks }}\\
\vskip 25pt
{ Sacha Davidson $^{1,}$\footnote{s.davidson@ipnl.in2p3.fr} and
S\'ebastien Descotes-Genon $^{2,}$}\footnote{descotes@th.u-psud.fr} 
 
\vskip 10pt  
$^1${\it IPNL, Universit\'e de Lyon, Universit\'e Lyon 1, CNRS/IN2P3, 4 rue E. Fermi 69622 Villeurbanne cedex, France}\\
$^2${\it Laboratoire de Physique Th\'eorique,
CNRS/Univ. Paris-Sud 11 (UM8627), 91405 Orsay Cedex.
 } \\
\vskip 20pt
{\bf Abstract}
\end{center}
\begin{quotation}
  {\noindent\small 
Scalar leptoquarks, with baryon and lepton number conserving interactions,
could have TeV scale masses, and be produced at  colliders or 
contribute to a wide variety of rare decays. In pursuit of some insight as to the
most sensitive search channels, we  assume that the
leptoquark-lepton-quark coupling can be constructed from
the known mass matrices. We estimate  the rates for selected
rare processes in three cases: leptoquarks carrying
lepton and quark flavour, leptoquarks with quark flavour only,
and unflavoured leptoquarks. We  find that leptoquark
decay to top quarks is an interesting search channel.
\vskip 10pt

}

\end{quotation}

\vskip 20pt  

\setcounter{footnote}{0}
\renewcommand{\thefootnote}{\arabic{footnote}}

\section{Introduction}
\label{intro}

Like  Higgs bosons, scalar leptoquarks $S$  have renormalisable
couplings $ \lambda^{ LQ}$ to two fermions. One could anticipate that 
the flavour structure of the leptoquark  and of the Yukawa couplings
 has the same
origin, suggesting that, from a phenomenological bottom-up
perspective,  the   $ \lambda^{ LQ}$ might be constructible
out of the Standard Model (SM) Yukawa matrices $Y_f$. 
We explore various possibilities
in this paper. Since our
building blocks are the known mass matrices,
we call this ``Minimal Flavour Violation (MFV) for leptoquarks''.

Leptoquarks \cite{LQrev} do not address  topical open questions
such as the identity of dark matter or the
origin of the electroweak scale. However, they
have some motivation.
The SM contains bosons with either  colour  (gluons)
or charge (Higgs and SU(2)  gauge bosons),
but no coloured and charged bosons which
could have  renormalisable interactions with a lepton
and a quark (leptoquarks). Nonetheless,    anomaly
cancellation implies that the quark and lepton sectors
are related. Theories (GUTs)  that unify the strong
and electroweak interactions frequently have B and L
violating   leptoquark
gauge bosons, { whose values are kept} 
 at the GUT scale because they could mediate proton decay\cite{pdec}. 
In this paper, we are interested in B and L
conserving scalar leptoquarks, with LHC-accessible masses. 
They could  arise in  technicolour models
\cite{Farhi:1980xs},
in  R-parity violating Supersymmetry (see {\it e.g.} \cite{Barbier:2004ez}), or
 be low-energy remnants of a
GUT\cite{Gershtein:1999gp}.
From a more phenomenological perspective,
  their  strong interactions
make them interesting for hadron colliders, and their
lepton-quark couplings  can be probed in
rare decay experiments. The prospects for detecting leptoquarks
at the LHC have been discussed in various models
\cite{BG,models}.

Some insight into ``where to look
for leptoquarks'', would be useful.
For instance, are B decays,
involving  third generation fermions, a more
promising place to look for leptoquarks than the more
sensitive Kaon decays?  
{ And how do they compare with $t\bar{t}$
production at hadron colliders ?}
To address such questions 
 requires  some knowledge about
the leptoquark masses $m_{S}$ and couplings $\lambda^{LQ}$.
From a  phenomenological perspective, 
the first solid step is to extract from
current  rare process data the bounds on  $|\lambda^{LQ}|^2/m_S^2$,
as was done recently, for instance, in
\cite{ybook,PRD,Carpentier}.  In this paper,
we fix $m_{S} \sim 300 $ GeV, which is
of order of the Tevatron lower limits\footnote{The Tevatron
bounds vary depending on the final state
fermions  and assumed Branching Ratios.} \cite{tevatron}
\beq
m_S \gsim  250-300~ {\rm GeV}
\label{tev}
\eeq
and accessible to the LHC \cite{coll}. Then we explore
various patterns for the $\lambda^{LQ}$,
 constructed  from the SM Yukawa couplings, which 
are consistent with the  bounds.
The patterns give predictions for  the most
promising search channels for leptoquarks.

To combine SM Yukawa couplings into a leptoquark coupling is
not immediately obvious, because
the quark  Yukawa couplings  connect  
the quark flavour spaces, 
and   $Y_e$ connects the lepton flavour spaces:
\bea
u_R \leftarrow Y_u \rightarrow q \leftarrow Y_d \rightarrow d_R  \nonumber \\
~\nonumber  \\ 
~~~~~~~~~~~~~~~~~~~~~~~~~~~ \ell \leftarrow Y_e \rightarrow e_R 
\nonumber 
\eea
but there is no bridge  between leptons and quarks. 
We consider three independent ways to construct a lepton-quark-leptoquark
interaction, which differ by the  quark- and lepton-flavour  assigned 
to the leptoquarks: 1)~flavour-singlet leptoquarks; 
2)~leptoquarks with non-trivial transformation properties 
only under the quark-flavour groups; 
3)~leptoquarks with both quark and 
lepton flavour indices.
In the first case, we are forced to introduce a new 
flavour-breaking structure connecting the quark and lepton flavour spaces.
In the second, following \cite{NS},
we can build flavour-invariant operators 
using the majorana neutrino mass matrix,
and in the third case, we need only the SM Yukawa couplings as symmetry
breaking structures. In all cases, there are a number of options,
so we make additional assumptions, aimed  to maximise the leptoquark
coupling $\lambda^{LQ}$.

\section{Notation}
\label{notn}

\subsection{Flavour symmetries and symmetry breaking terms}

The kinetic terms of three generations of  Standard Model 
(SM)  fermions have a global
$U(3)_q \times U(3)_u \times   U(3)_d \times U(3)_\ell \times U(3)_e $
symmetry, which is broken  by the quark and
charged lepton Yukawa couplings to $U(1)_Y
\times U(1)_B \times U(1)_{Le} \times U(1)_{L\mu}
\times U(1)_{L \tau}$ (where $U(1)_Y$ is global hypercharge,
 $U(1)_B$ is baryon number, and the $ U(1)_{Li}$ are lepton flavours).
Majorana neutrino masses\footnote{Dirac 
neutrino masses are beyond the scope of this preliminary analysis.
They allow additional leptoquarks,  
interacting with  the singlet $\nu_R$s. But larger
quark and lepton flavour-changing rates might be obtained
in this case.}
 will be included in section \ref{qflav}.   As is
well known,   the Yukawa couplings give hierarchical
masses to the charged leptons and quarks,
and identify a unique ``mass eigenstate basis'' in the
flavour spaces of the $u,d,e$ and $\ell$.
In the flavour space of the $q$s,
 there are the two mass eigenstate
bases of the $d_L$s and the $u_L$s, 
which are related by the CKM matrix $K$. 

There is a large body of
precise flavour data  in the quark sector,
which agrees with SM expectations. This implies
that any flavoured interactions of
new particles accessible to the LHC,
should, somehow, ``align'' themselves on the
eigenbases of the quark Yukawa couplings and share their
eigenvalues.  This is elegantly
obtained with the Minimal Flavour Violation
(MFV) hypothesis \cite{MFV,dAGIS}: only the
Yukawa matrices  can break the global $U(3)^5$. 
Extending the MFV framework to 
sectors  other than the quarks  has been
previously considered by various authors\cite{MFVext}.

We write the SM  Yukawa matrices with the flavour indices ordered
doublet-singlet. For instance, in the case of the charged leptons:
\beq
[Y_e]^{In}  \langle H \rangle  \overline{e}_n \ell_I + h.c.~~
\supset  ~~ m_e \overline{e_R} e_L + h.c. 
\eeq
where $\langle H \rangle = v = $  175 GeV and
the diagonal Yukawa matrix of fermion $f$ will be denoted
$D_f$.  Capitalised roman indices
 $I,J,K$  correspond to SU(2) doublets  $q, \ell$, lower case 
roman indices  ($i,j,k$)  are carried by the singlets: $e,u,d$. 
Preference is given to the beginning of the alphabet for leptons,
and the later part for quarks. Chiral subscripts $L,R$ are
suppressed to avoid confusion with flavour indices.

We  take the perspective that the
 largest
eigenvalue  of the 
$[Y_f]$ may be ${\cal O}(1)$. 
We implement this by
 considering a Higgs sector of
two  doublets,  $H_u$ and $H_d$,
coupled separately to up-type quarks 
($H_u$)
and down-type quarks and charged leptons ($H_d$). This possibility 
allows us to change the relative normalization of the Yukawa 
couplings, changing the ratio of the two Higgs 
vacuum expectation values: 
$\tan \beta = \langle H_u \rangle/ \langle H_d \rangle$. 
In particular, for $\tan \beta \gg 1$, we have:
\beq
D_e \equiv {\rm diag} \{ y_{e}, y_{\mu} , y_\tau \} 
= \frac{\tan \beta}{v} {\rm diag} \{ m_{e}, m_{\mu} , m_\tau \}~,   
D_d = \frac{\tan \beta}{v} {\rm diag}\{m_d, m_s,m_d\}~,  
D_u = \frac{1}{v} {\rm diag}\{m_u, m_c,m_t\}~,
\label{deftb}
\eeq
and the one-Higgs doublet case is recovered for $\tan \beta = 1$. 
Given the misalignment of the two quark Yukawa couplings is not 
affected by their overall normalization, present flavour 
data are compatible with large $\tan\beta$ values. The latter 
choice is particularly interesting since it could allow to 
consider a scenario where top, bottom and tau Yukawa couplings 
are of   order~1.

For further convenience, we define the following 
combinations of SM Yukawa matrices:
\beq
W_e = Y_e Y_e^\dagger~~~,~~~
W_d = Y_d Y_d^\dagger~~~,~~~
W_u = Y_u Y_u^\dagger~~~,~~~
\widetilde{W}_e = Y_e^\dagger Y_e~~~,~~~
\widetilde{W}_d = Y_d^\dagger Y_d~~~,~~~
\widetilde{W}_u = Y_u^\dagger Y_u~~~.~~~
\eeq

In section \ref{qflav}, we will include neutrino masses, 
assumed to be majorana, with a mass matrix $[m_\nu]$ included
in the Lagrangian as
\beq
\frac{[m_\nu]_{IJ}}{2} \overline{\nu^c}_I \nu_J + h.c. 
\eeq

\subsection{Leptoquarks}

We consider 
SU(2) singlet and  doublet  scalar leptoquarks, 
with renormalisable $B$ and $L$
conserving interactions. 
In the notation of
Buchmuller,Ruckl and Wyler\cite{BRW} 
\footnote{We took the complex conjugate of ${\cal L}$, to obtain
fermion field order lepton-quark, without
taking the  hermitian conjugate of the $\lambda$s. So our $\lambda$s are
$\lambda_{BRW}^*$.}, these
can be added to the SM Lagrangian  as:
\bea
{\cal L}_{LQ} & = & 
S_0 ( {\bf \lambda_{L S_0}} \overline{\ell} i \tau_2 q^c
+ {\bf \lambda_{R S_0}} \overline{e} u^c ) +
\tilde{S}_0 {\bf \tilde{\lambda}_{R \tilde{S}_0}} 
\overline{e} d^c   
+
 ( {\bf \lambda_{L S_2}} \overline{\ell}  u
+ {\bf \lambda_{R S_2}}  
\overline{e } q [i \tau_2  ])S_{2} +
 {\bf \tilde{\lambda}_{L \tilde{S}_2}} 
\overline{\ell} d \tilde{S}_2
+ h.c.
\label{BRW}
\eea
where the $\lambda$s are 3 $\times$ 3 matrices 
with index order lepton-quark, and  $\tau_2$ is a Pauli
matrix, so $i \tau_2$ provides the antisymmetric SU(2) contraction. 
Notice that 
$ ( \overline{q^c}_P i\tau_2 \ell_I) \sim u_Pe_I - d_P \nu_I$,
so the leptoquark 
$S_0$ does not interact with $de$, and cannot mediate
processes such as $K_L \to \mu e$  at tree level (see
table \ref{4fv}).
In this Lagrangian, the leptoquark leaves the vertex into
which enter  the leptons. 
So  for instance, the SU(2) singlets leptoquarks have fermion number  2,  
the doublets carry no fermion number. Other quantum
numbers are listed below
($Q_{em} = Y/2 + T_3$, $T_3 = \pm 1/2$ for doublets)
\beq
\begin{array}{lccccc}
{\rm leptoquark}  & Y  &B & L  & SU(2) & {\rm couplings}  \\ 
S_0 & -2/3 & 1/3 & 1 & 1 &  {\bf \lambda_{L S_0}},  {\bf \lambda_{R S_0}}\\
\tilde{S}_0 & -8/3 & 1/3 & 1 & 1 & {\bf \tilde{\lambda}_{R \tilde{S}_0}}   \\
S_{2} & -7/3 & -1/3 & 1 & 2 &  {\bf \lambda_{L S_2}} ,  {\bf \lambda_{R S_2}} \\ 
\tilde{S}_2 & -1/3 & -1/3 & 1 & 2 &  {\bf \tilde{\lambda}_{L \tilde{S}_2}} 
\end{array} 
\eeq

In addition to the Tevatron lower bound on leptoquark masses
given in eq. (\ref{tev}),
there is a  lower bound from HERA\cite{HERA} $m_S > 250-300$ GeV 
{ (for leptoquarks coupling to first-generation fermions with $\lambda\sim 0.1$)}
and a variety of
constraints from low energy/precision
experiments which are sensitive to
the  coefficients of  dimension six operators
mediated by  leptoquarks. Such operators include the
quark and charged lepton dipoles
 and  four fermion operators
involving a  quark, an anti-quark, a lepton and  an anti-lepton (which
we refer to as two-quark-two-lepton operators). 
The four-fermion
operators can be Fierz-rearranged to  
(lepton-current)$\times$ (quark current) form  (see table
\ref{4fv}) which is more
convenient for comparing to SM processes. 
Following  \cite{ybook}, the coefficients of
these $V\pm A$ two-quark-two-lepton operators
can be normalised as
\beq
\frac{C_X^{ijrs}}{m_{S}^2} = \varepsilon^{ijrs}_X \frac{4 G_F}{\sqrt{2}} 
\label{defve}
\eeq
and experimental constraints can be set on 
these four-index $\varepsilon^{ijpq}$s .We
 use the recent  bounds of \cite{Carpentier},
which arise  from
 leptonic 
and semi-leptonic decays of pseudoscalar
mesons ({\it e.g.} 
$R_K \equiv \Gamma(K^+ \to e^+ \nu)/\Gamma(K^+ \to \mu^+ \nu)$ 
and $K^+ \to \pi^+ \nu \bar{\nu}$), 
flavour-changing but
generation-diagonal meson decays such as
$K_L \to \mu^\pm e^\mp$, contact interaction searches at colliders
and   $\mu -e$ conversion on nuclei (we neglect leptoquark loop
contributions to four-quark operators,
which are constrained by meson-anti-meson mixing \cite{PRD}).
{ Considering absolute values only} 
and assuming $m_S \sim  300$ GeV,
 the upper bounds of  \cite{Carpentier} on
the  $\varepsilon$s imply
that 
\beq
\label{6}
\frac{\lambda^2}{6} \lsim  \varepsilon
\eeq for appropriate indices.

\begin{table}[htb]
\begin{center}
$
\begin{array}{||l|l|l||}\hline
{\rm interaction} & {\rm 4-fermion ~  vertex} & {\rm
Fierz-transformed ~ vertex} \\ \hline
%
%
( \lambda_{LS_0} \overline{q^c} i\sigma_2 \ell +
 \lambda_{RS_0} \overline{u^c} e) S_0^{\dagger} 
&
 \frac{\lambda_{RS_0}\lambda_{RS_0}^*}{m_S^2} ( \overline{u^c} e)(\bar{e} u^c) 
&
 \frac{\lambda_{RS_0}\lambda_{RS_0}^*}{2m_S^2} ( \bar{u} \gamma^{\mu}P_R u)
    (\bar{e} \gamma_{\mu} P_R e)  \\
& \frac{\lambda_{LS_0}\lambda_{LS_0}^*}{m_S^2}
 ( \overline{q^c}  i\sigma_2 \ell)(\bar{\ell} i\sigma_2 q^c) &
\frac{\lambda_{LS_0}\lambda_{LS_0}^*}{2m_S^2} 
 (  \overline{u}  \gamma^{\mu}P_L u )(\bar{e}
\gamma_{\mu}P_L e) \\
& 
&
\frac{\lambda_{LS_0} \lambda_{LS_0}^*}{2m_S^2} (  \bar{d}  \gamma^{\mu} P_L u )(\bar{\nu}
\gamma_{\mu}  P_L e) \\
&  
&
\frac{\lambda_{LS_0} \lambda_{LS_0}^*}{2m_S^2} ( \bar{d}  \gamma^{\mu}P_L  d)(\bar{\nu}
\gamma_{\mu} P_L \nu) \\
& 
\frac{\lambda_{RS_0} \lambda_{LS_0}^*}{m_o^2}
 ( \bar{q}^c i\sigma_2 \ell)(\bar{e}
 u^c) & \frac{\lambda_{RS_0} \lambda_{LS_0}^*}{2m_S^2} ( \bar{u} P_Lu )
 (\bar{e} P_L e)  +...\\
&   
& 
\frac{\lambda_{RS_0} \lambda_{LS_0}^*}{2m_S^2} ( \bar{u}  P_Ld)(\bar{e}  P_L  \nu) +...
\\ \hline
%
%
\lambda_{R\tilde{S_0}}  \overline{d^c} e \tilde{S}_0^{\dagger} 
&
 \frac{\lambda_{R \tilde{S_0}}\lambda_{R \tilde{S_0}}^*}{{m}_S^2}
 (  \overline{d^c} e)(\bar{e} d^c) 
& 
\frac{\lambda_{R\tilde{S_0}}\lambda_{R\tilde{S_0}}^*}
{2{m}_S^2} (  \overline{d}
\gamma^{\mu}  P_R d)(\bar{e} \gamma_{\mu} P_R e)  
\\ \hline
%
%
(\lambda_{LS_{2}} \bar{u} \ell + \lambda_{RS_{2}}
\bar{q} i\sigma_2   e) S_{2}^{\dagger} 
&
 \frac{\lambda_{LS_{2}}\lambda_{LS_{2}}^*}{m_S^2} ( \bar{u} \ell)(\bar{\ell} u) 
&
-\frac{\lambda_{LS_{2}} \lambda_{LS_{2}}^*}{2m_S^2} ( \bar{u} \gamma^{\mu} P_R
u)(\bar{\nu} \gamma_{\mu}P_L \nu)  \\
& 
&
-\frac{\lambda_{LS_2} \lambda_{LS_{2}}^*}{2m_S^2} ( \bar{u} \gamma^{\mu} P_Ru)
(\bar{e} \gamma_{\mu}P_L  e) \\
& 
\frac{\lambda_{RS_2}\lambda_{RS_2}^*}{m_S^2} ( \bar{q} e)(\bar{e} q) 
&
-\frac{\lambda_{RS_{2}}\lambda_{RS_2}^* }{2m_S^2} ( \bar{u} \gamma^{\mu} P_L u)
(\bar{e} \gamma_{\mu} P_R e) \\
&  
&
-\frac{\lambda_{RS_{2}} \lambda_{RS_2}^*}{2m_S^2} ( \bar{d}_L \gamma^{\mu}P_L d)
(\bar{e} \gamma_{\mu}P_R e) \\
& 
\frac{\lambda_{LS_{2}} \lambda_{RS_{2}}^*}{m_S^2}
 ( \bar{u} \ell)(\bar{e} q) 
&
-\frac{\lambda_{LS_{2}} \lambda_{RS_{2}}^*}{2m_S^2} ( \bar{u} P_L d)
(\bar{e} P_L \nu) +..\\
& &
-\frac{\lambda_{LS_{2}} \lambda_{RS_{2}}^*}{2m_S^2} ( \bar{u} P_L u)
(\bar{e} P_L e) +..
\\ \hline
%
%
\lambda_{L \tilde{S}_{2}} \bar{d} \ell  \tilde{S}_{2}^{\dagger}
 & 
\frac{ \lambda_{L\tilde{S}_{2}} \lambda_{L\tilde{S}_{2}}^*}{\tilde{m_S}^2} (
 \bar{d} \ell)(\bar{\ell} d) 
&
-\frac{ \lambda_{L \tilde{S}_{2}} \lambda_{L \tilde{S}_{2}}^*}{2\tilde{m_S}^2} (
\bar{d} \gamma^{\mu}P_R d)(\bar{\nu} \gamma_{\mu}P_L  \nu)  \\
& 
&
-\frac{ \lambda_{L \tilde{S}_{2}} \lambda_{L \tilde{S}_{2}}^* }{2\tilde{m_S}^2}
( \bar{d} \gamma^{\mu} P_R d)
 (\bar{e} \gamma_{\mu} P_L e)  
\\  \hline
\end{array} $
\end{center}
\caption{Four fermion operators induced by leptoquarks. After Fierz
rearrangement, the effective interactions $\propto \lambda_L \lambda_R$
also have tensor components, represented as $+...$, which we neglect.}
\label{4fv}
\end{table}

At dimension six, there are also 
  flavour changing  dipole operators  involving two fermions
 and a gauge boson ($\gamma, Z, g $), which can contribute
to anomalous $Z$ decays \cite{Concha} 
and processes such as $b \to s \gamma$ and
$\mu \to e \gamma$.
After Spontaneous Symmetry Breaking, the dipole
that mediates $f_2 \to f_1 \gamma$ can be written as
 \cite{ybook}
\beq
\frac{e_{em} m_2}{2}   
\overline{f}_1 \sigma^{\a \b}( A_L P_L + A_RP_R )f_2 F_{\a \b} + h.c. 
\equiv \frac{e_{em} m_2 G_F }{2} 
\overline{f}_1 \sigma^{\a \b}(  \varepsilon^{f_1f_2}_L P_L + 
\varepsilon^{f_1f_2}_RP_R )f_2 F_{\a \b}  + h.c.
\label{dipop}
\eeq
where the second equality defines the dimensionless
(two-index) $\varepsilon$s for these operators. Notice that  
these effective couplings are  defined factorising
the heavy fermion mass, as expected from a
chirality flip on the  external leg.
 They can be bounded from the experimental limits
on  the branching ratio, for instance as 
\beq
BR(\mu \to e \gamma)
=
\frac{48 \pi^3 \alpha}{G_F^2} \left( A_L^2 + A_R^2 \right)
= 48 \pi^3 \alpha \left( |\varepsilon^{e \mu}_L|^2 + 
|\varepsilon^{e \mu}_R|^2 \right)
\label{Gtmg}
\eeq
A  list of experimental  bounds  on selected
$\varepsilon$s, relevant to the
scenarios we consider,  can be found in   tables 
\ref{results3}, \ref{results4} and \ref{results5}.

\begin{figure}[ht]
\begin{center}
\epsfig{file=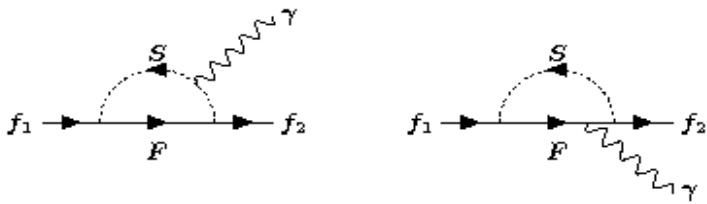, height=3.5cm,width=10cm}
\end{center}
\vspace{-10mm}
\caption{One-loop diagrams mediated  by a leptoquark $S$
that could induce $f_1 \to f_2 \gamma$.}
\label{figf1f2g}
\end{figure}

We estimate  $\varepsilon^{f_2 f_1}$,
arising from the diagrams of figure \ref{figf1f2g}
and of wave function renormalisation, as \cite{Lavoura}
\beq
\varepsilon^{f_2 f_1} = \frac{1}{96 \pi^2} \times
\left\{
\begin{array}{ll}
\lambda_X \lambda_X^* (Q_F + Q_S/2) &  \\
\frac{m_F}{m_{1}}
\lambda_L \lambda_R^* ( [9  +  6 \ln (\frac{m_F^2}{m_S^2} )] 
Q_F - 3 Q_S)  & \\
\end{array}
\right.
\label{f1f2g}
\eeq
where $X = L$ or $R$, and the sign of the electric charges
$Q$ is given by the line directions of figure \ref{figf1f2g}.
These estimates apply to the case $m_F \to 0$; 
for  a top quark in the loop,  the
numerical factor is a bit smaller \cite{Lavoura}.

In principle, additional constraints on 
leptoquarks  could be obtained from
electroweak precision observables  (such
as the oblique parameters $S,T,U$...). 
However, by construction,
the oblique parameters are sensitive to
the breaking of the SM gauge symmetry
(the relevant non-renormalisable
operators contain the Higgs field),
and  the Higgs-leptoquark
couplings~\cite{Hirsch:1996qy} do not
concern  our analysis.  Therefore we expect electroweak
precision contraints  on the flavoured
leptoquark couplings to be
unimportant.
To check this, one can  estimate  \cite{KMa}
\bea
S  &  \equiv &
\frac{16 \pi^2c_W}{g^2s_W} 
\frac{d}{dq^2} \Pi_{W_3 B}(q^2) {\Big |}_{q^2 = 0} 
\simeq  -N_S N_c \frac{Y}{6 \pi} \frac{\Delta^2}{m_S^2}
\\
T & \equiv & \frac{1}{\alpha} \left( \frac{\Pi_{WW}(0)}{m_W^2} - 
\frac{\Pi_{ZZ}(0)}{m_Z^2}  \right)\simeq 
\frac{ N_SN_c }{16 \pi s^2_W  m_W^2} \Delta^2
\eea
where
 $\Delta^2$ can be  the mass-squared splitting in a doublet,
or the  singlet-doublet mixing mass $m_S \Delta S_0 \tilde{S}_2$
\cite{Lavoura:1993mz},
and $N_S$ is the number  of copies  of the leptoquark:
three if the leptoquark has quark flavour, nine if it
has quark and lepton flavours.  
Requiring $0 \lsim T-S \lsim .1$, $-.25 \lsim S+T \lsim .25$
\cite{PDB}
suggests that 
\beq
N_S \frac{\Delta^2}{m_S^2} \ll 1
\label{STU}
\eeq
is acceptable. A Higgs-leptoquark coupling will neccessarily
arise in our models at one loop.   If the mass splitting is induced by
a third generation fermion loop, with the Higgs boson  interacting
with the $t$ quark, then $\Delta^2 \lsim v^2/16 \pi$, 
eq. (\ref{STU}) is satisfied even for 9 leptoquark
flavours, and we conclude the oblique parameters
do not provide significant constraints on the masses
and flavoured couplings of leptoquarks.

\section{Flavour singlet leptoquarks}
\label{unit}

Ideally, we would like to construct, out of SM Yukawa matrices,  
a leptoquark coupling matrix  $[\lambda]$,  with one quark flavour index
and one lepton flavour index. To see why this is not possible,
recall the
 popular formulation of MFV \cite{dAGIS}, which
identifies the Yukawa matrices as
auxiliary fields, or ``spurions'',
whose   transformations  
under the global $U(3)^5$ are
chosen to ensure  the invariance of  
the Yukawa interactions. 
Then the spurions get ``vacuum expectation
values'', and the $U(3)^5$ is ``spontaneously
broken'', leaving baryon and lepton number
as global symmetries.  That is, Lorentz scalars constructed out
of SM fermions and spurions do not
transform under $B$ or $L$. However, 
since the leptoquarks carry baryon
and lepton number, 
$[\lambda]^{LQ}$ (where $L$ and $Q$ are 
SM lepton  and quark fields)
 cannot be constructed out of
the SM spurions.
 So  in this section we consider  adding a new spurion.

\subsection{Adding a new spurion}
\label{add}

This new interaction should  depart as little
as possible from SM  flavour structures,  and should be
consistent with current bounds. 
We therefore make three assumptions. 
First, we  connect   the $u$  singlets to leptons,
because  constraints on new
interactions  involving $u,c,t$ quarks
are weaker than those involving $d$-type quarks.
Secondly, the spurion is taken to
be proportional to the  unit matrix ${\cal I}$, because it only
breaks the $SU(3)_l \times SU(3)_u$ symmetry of the kinetic
terms to $SU(3)_{l+u}$ (where $l = \ell$ or $e$).
This suggests two possible  couplings:
\bea
 [{\bf \lambda_{R S_0}}]^{ip} S_0 \overline{e}_i u^c_p 
& =&
  \lambda_{S_0} [{\bf  {\cal I}}] ^{ ip}  S_0    \overline{e}_i u^c_{p} 
\label{eu} \\
{ [{\bf \lambda_{L S_2}} ]} ^{Ip}\overline{\ell}_I  u_p S_{2} 
& =&
 \lambda_{S_2} [{\bf  {\cal I}}]^{Ip} \overline{\ell}_I  u_p S_{2} 
\label{lu}
\eea
where the $\lambda_{S_0}$ and $\lambda_{S_2}$ are constants not
matrices.
If  $ [{\bf \lambda_{R S_0}}] \propto {\cal I}$
in  generic bases for the $u$ and $e$
flavour spaces, then  it would become  a unitary matrix in
the mass eigenstate bases of $e$ and $u$. This brings us to
our third assumption: we impose that the unit matrix is in
the mass bases(because otherwise there are severe constraints from
$\mu N \to e N$ and $\tau \to \pi^0 \ell$). 
 With these assumptions, the most restrictive
experimental bound  comes from Tevatron searches for
contact interactions of the form  
$(\overline{u} \gamma P_{R} u)
(\overline{e}\gamma P_{L,R} e)$, and gives
 $ \varepsilon_{eu}^{1111} <  10^{-2}$
and 
 $ \varepsilon_{\ell u}^{1111} < 1.4 \times 10^{-2}$.
 These would be satisfied,  for 
$m_{S} \sim 300$ GeV, by
 $\lambda_{S_0} ,  \lambda_{S_2} \lsim 1/4$.

From the  new spurion of eq. (\ref{eu}), we can 
construct  couplings  of leptoquarks carrying two
units of fermion number,
to  other types of
leptons or quarks by multiplying by the SM Yukawas:
\bea 
[{\bf \lambda_{L S_0}}]  S_0   \overline{\ell} i \tau_2 q^c
&  =  & 
    [{\bf  Y^*_e  {\cal I}  Y_u^\dagger  }]^{IP} 
S_0 \overline{\ell}_I i \tau_2 q^c_P
\label{e17}
\\
{ [{\bf \tilde{\lambda}_{R \tilde{S}_0}}]} \tilde{S}_0  \overline{e} d^c  
 & = & 
 [{\bf  {\cal I}  Y_u^{\dagger}  Y_d   }]^{ip}   
\tilde{S}_0 \overline{e}_i d^c_p
\nonumber
\eea
We have not included $\lambda_{S_0}$ in the definition
of these other couplings, because this allows them to be
larger, and because such an overall scaling of 
flavoured interactions may be due to some unflavoured
physics. So we identify as our new ''spurion''
the unit matrix ${\cal I}$, rather than 
 $\lambda_{S_0}{\cal I}$  or $\lambda_{S_2}{\cal I}$.
See also the comments  in section \ref{commentaires}.

Similarly, from the  new spurion of eq. (\ref{lu}), we can 
construct  couplings  of the  leptoquarks with zero fermion
number:
\bea 
{  [{\bf \lambda_{R S_2}}]} \overline{e } [ i \tau_2 q ]^T  S_2
&  =  & 
   [{\bf  Y^T_e  {\cal I} Y_u^T  }]^{iP} 
\overline{e}_i [ i \tau_2 q_P]^T  S_2 
\label{e18}
\\
{ [{\bf \tilde{\lambda}_{L \tilde{S}_2}}]}^{Ip}  
\overline{\ell}_I  d_p \tilde{S}_2
 & = &  
    [{\bf  {\cal I}  Y_u^{T}  Y^*_d   }]^{Ip}   
\overline{\ell}_I d_p\tilde{S}_2  ~~.
\nonumber
\eea

The   four fermion operators induced respectively by
$S_0$, $\tilde{S}_0$, $S_{2}$,  and $\tilde{S}_{2}$,
and their coefficients, are:
\begin{eqnarray}
& \frac{|\lambda_{S_0}|^2  }{m_S^2} 
( \overline{u^c}_i e_i)(\bar{e}_j u^c_j)
~~,~~~ 
 {[{\bf  D_e    Y_u^T  }]}^{IP}
  \frac{|\lambda_{S_0}|}{m_S^2}
 ( \overline{q^c}_P i\tau_2 \ell_I)(\bar{e}_j
 u^c_j) 
~~,~~
{ [{\bf  D_e    Y_u^T  }]}^{IP} 
{ [{\bf  D_e    Y_u^\dagger  }]}^{JS} 
 \frac{1}{m_S^2}
 ( \overline{q^c}_P  i\tau_2 \ell_I)(\bar{\ell}_J i\tau_2 q^c_S) & {  S_0}
\label{opS0}\\
& { [{\bf    D_u K  D_d   }]}^{ip}   
{ [{\bf   D_u K^*  D_d   }]}^{js}   
 \frac{1}{{m}_S^2}
 (  \overline{d^c}_p e_i)(\bar{e}_j d_s^c) & {\tilde{S}_0}
\label{optS0} 
\\
& \frac{|\lambda_{S_2}|^2}{m_S^2} ( \bar{u}_I \ell_I)(\bar{\ell}_J u_J)
~~~,~~ 
   [{\bf  D_e   Y_u^T  }]^{jS} 
\frac{|\lambda_{S_2}|}{m_{S}^2}
 ( \bar{u}_I \ell_I)(\bar{e}_j q_S) 
~~,~~
 { [{\bf  D_e   Y_u^\dagger  }]}^{iP} 
 [{\bf  D_e Y_u^T  }]^{jS} 
\frac{1}{m_{S}^2} ( \bar{q}_P e_i)(\bar{e}_j q_S) & {  S_2}
\label{opS2}\\
& {  [{\bf  D_u K^*  D_d   }]}^{Ip} 
  [{\bf    D_u K  D_d   }]^{Js}   
\frac{1 }{{m_{S}}^2} (
 \bar{d}_p \ell_I)(\bar{\ell}_J d_s) & { \hfill \tilde{S}_2}
\label{optS2}
\end{eqnarray}
where the ${\bf Y^T_u} = {\bf D_u} $ [ $ = {\bf D_u K^*}$]
 for an up-type [down-type]  quark
on the external leg.
  Notice  that generation number
can only change when down-type quarks are involved.

The amplitudes induced by these operators are suppressed by
zero, two or four Yukawa eigenvalues. By construction,
the couplings $\propto |\lambda_{S_0}|^2$, $|\lambda_{ S_2}|^2$
of the operators unsuppressed by Yukawa couplings,  are  generation
diagonal, and  small enough to
satisfy the bounds. 

Now, consider the
  pseudoscalar operators (middle operator
of eq. (\ref{opS0}) and eq. (\ref{opS2})), which are suppressed
by $y_f^2$.  These operators, mediated by $S_0$ 
or $S_2$, always contain
an up-type quark and a   lepton of the same generation, so 
they can  induce $D_0 \to \mu^\pm e^\mp$,
but not $K_L\to  \mu^\pm e^\mp$ or 
$N \mu \to N e$.
 Only in the charged current
case, where the down-type quark brings an element  of
the CKM matrix $K$, is non-conservation of generation number possible.
These operators  contribute to the leptonic decay of 
pseudoscalar mesons $M^+ = \pi^+,K^+$ or $B^+$.  Recall that in the SM, 
the $V-A$  amplitude for $M^+$ decay
is   suppressed  by a factor of the charged lepton mass,
which is required to flip the chirality on the external leg.
This suppression is absent for a pseudoscalar
operator. 
The leptoquark $S_0$   
induces $M^+ \to  \nu_\tau \bar{e}$,
with $\varepsilon \sim |\lambda_{S_0}| K_{tm} y_t y_\tau/6$,
where $m= d,s,b$ as the case may be. This expectation
 is less than the
experimental bound for
$\pi$ and $K$ decays, which is
  $\varepsilon \lsim 2 \times 10^{-5}$ \cite{Carpentier}.
The decay $B^+ \to \nu e^+$ is not observed,
consistently with its tiny expected  SM branching ratio.
However, we can compare to the SM prediction for the observed
decay $B^+ \to \tau^+ \nu$:
\beq
\frac{BR(B^+ \to e^+ \nu_\tau)}{BR(B^+ \to \tau^+ \nu)}
\simeq
\frac{m_B^2}{m_\tau^2}
\frac{ |\lambda_{S_0} K_{tb} y_t y_\tau|^2}{|K_{ub}|^2} \sim 870 y_\tau^2
\sim 870 \frac{m_\tau^2}{v^2} \tan ^2 \beta
\label{Bnue}
\eeq
For $\lambda_{S_0} \simeq 1/4$ and  $\tan \beta \gsim 1/\sqrt{2}$,
this exceeds the upper bound $BR(B^+ \to e^+ \nu) \lsim 4 \times 10^{-2}$
$BR(B^+ \to \tau^+ \nu)$ by a factor $\sim 2$. So for $\tan \beta = 1$,
$B^+\to e^+ \nu$ provides the best bound on $S_0$  with both
chiral couplings, in this model. 
Or, if we allow $\tan \beta$ as a free parameter,  
 $B^+ \to e^+ \nu$ is a sensitive probe for this pattern of
couplings for the singlet $S_0$. 

Charged current pseudoscalar operators 
 are also induced by the  doublet leptoquark
$S_2$, but with different index contractions, such that
it induces  the observed  $B^+ \to  \nu_e \bar{\tau}$
with  $\varepsilon \sim |\lambda_{S_2}| K_{tb} y_t y_\tau/6$.
This contribution can be competitive with the
SM, for $\tan \beta\gsim 4$, which can be interesting in view
of the experimental anomaly in 
$BR(B^\pm \to \tau ^\pm \nu)$ (see {\it e.g.} \cite{Btaun}).

 It is easy to check  that 
the operators suppressed by    the fouth power
of Yukawa eigenvalues  are
of $V \pm A $ form, and  harmless.
For instance, experimentally, 
 $K_L \to \bar{\mu}e$  gives one of the
most stringent  limits on  generation-diagonal
flavour violation:
 $\varepsilon \lsim  3 \times 10^{-7}$.  The
expectation from $\tilde{S}_2$ or $\tilde{S}_0$ exchange   would be
$\varepsilon \sim  y_u K_{ud} y_d y_c K_{cs} y_s/6 
\sim  2 \tan ^2 \beta  \times 10^{-16}$.
Generation number changing
operators  proportional to $y_f^4$, 
neccessarily  involve down-type quarks, and
are CKM suppressed. For instance for $\mec$,
 $\varepsilon \sim | y_u y_d  y_c K_{cd} y_d|/6$ 
 at tree level, which is unobservably small  even
for  large $\tan \beta$. The
expectation for $B_s \to \mu \bar{\mu}$
(mediated by $\tilde{S}_0$ or  $\tilde{S}_2$) 
would be  $\varepsilon \sim | y_c K_{cb} y_b y_c K_{cs} y_s|/6
$, which is much less than the experimental limit \cite{Carpentier}
of $ 7 \times 10^{-5}$.
Since lepton generation change  occurs via CKM mixing angles,
which appear in $\lambda$ suppressed by  quark Yukawa eigenvalues,  
$\tmg$ is more sensitive than $\meg$ because it involves higher
generations.  
The  $\tmg$
loop has an internal down-type quark, and 
 gives \cite{Lavoura} (for  $\tilde{S}_0$ ---
the factor of 1/6 is a function of the charges of the loop
particles):
\beq
\varepsilon \sim  \frac{1}{6} \frac{N_c}{16 \pi^2}
 y_t  y_c \sum_r K_{tr} y_{d,r}^2 K_{c r}
 \sim  \frac{1}{96 \pi}
  y_c  y_{b}^2 K_{c b}
\label{to1}
\eeq
which is less than the experimental bound (see table \ref{results3}).
In the last approximation of eq. (\ref{to1}),  and in
all the tables, we
approximate $\pi \simeq N_c$, and  $y_t \simeq K_{ii} \simeq 1$.
The most sensitive process we found was 
$ K^+\to \pi^+ \nu_\tau \bar{\nu}_\tau$ due to $S_0$ exchange,
for which we estimate 
\beq
6 \varepsilon \sim y_\tau ^2 y_t^2 K_{td} K_{ts} ~~,~~ {\Big (}
 y_t^2 K_{td} K_{ts} y_d y_s
{\Big )} < 5 \times 10^{-5} {\Big |}_{expt}
\eeq
which is in agreement with the bounds for $\tan \beta \lsim 45$
(in parentheses is the estimate  for $\tilde{S}_2$ exchange).

Table \ref{results3} lists some interesting rare  processes, with the
experimental bounds, and, in the third colomn, 
the largest rates mediated by any of the leptoquarks with this
pattern of couplings. 
These estimates suggest that our
``unflavoured'' leptoquarks 
${S}_0$ or  ${S}_2$ could  be found in leptonic  $B$ decays 
$B^+ \to l^+ \nu$, and that the rare  process  most sensitive
to $\tilde{S}_0$  is $\tmg$.

Alternatively, hadron colliders could search for
all these leptoquarks, which  would decay near the
production point as assumed in Tevatron searches. 
${S}_0$  and  ${S}_2$ decay with similar branching ratios
to $e u$, $\mu c$ and $\tau t$, and 
$\tilde{S}_0$ and  $\tilde{S}_2$ would preferentially
decay to third generation fermions.

\renewcommand{\arraystretch}{1.45}

\begin{table}
\begin{center}
$
\begin{array}{||l|l|l||}\hline
{\rm process} & \varepsilon < & {\rm  {estimates}~ (unflavoured ~S)} \\
\hline \hline
R_K   &2 \times 10^{-5} 
& \frac{1}{6}|\lambda_{S_0}| K_{ts} y_t y_\tau \sim {\rm 2 \times}10^{-5} t_\beta 
\\
%
&& 
\\
\hline
B^+ \to e^+\nu   & 2 \times 10^{-4}  
& \frac{1}{6} |\lambda_{S_0}|y_\tau y_t K_{tb}  \sim 3 \times 10^{-4} t_\beta \\  
 & &\\
\hline
B^+ \to \tau^+\nu & 8 \times 10^{-4}  
& \frac{1}{6} |\lambda_{S_2}|y_\tau y_t K_{tb} \sim 3 \times 10^{-4} t_\beta  
\\
 & &\\
\hline
D_0 \to \mu^\pm e^\mp  & 6 \times 10^{-4}  
& \frac{1}{6} |\lambda_{S_0}| y_\mu y_c \sim 2 \times 10^{-7} t_\beta   
 \\ 
  & & \\
\hline\hline
%
%
K^+ \to \pi^+ \nu \bar{\nu} & 9 \times 10^{-6}  
& \frac{1}{6} y_\tau^2 y_t^2 K_{ts} K_{td}  \sim 5 \times 10^{-9} t_\beta^2   
\\
  & &\\
\hline
K_L \to \mu^\pm e^\mp &3 \times 10^{-7} 
&  \frac{1}{6} y_uy_d y_c y_s 
\sim 2 \times 10^{-16} t_\beta^2   
\\
   & &\\
\hline
B_s \to \mu^\pm \mu^\mp & 7 \times 10^{-5} &  
\frac{1}{6} y^2_c K_{cb} y_b  y_s 
\sim 5 \times 10^{-12} t_\beta^2  
\\
   && \\
\hline
B^+ \to K^+ \tau^\pm \mu^\mp & 2 \times 10^{-3}  &   
\frac{1}{6} y_t y_b y_c y_s 
\sim  2 \times 10^{-8} t_\beta^2   
\\
   & &\\
\hline
\mu N \to e N' &8 \times 10^{-7} &
 \frac{1}{96 \pi^2}
 y_c  y_u  K_{cb} y_{b}^2 K_{u b}
\sim   10^{-17} t_\beta^2   
\\
   & &\\ 
\hline
\mu \to e \gamma  &  \sim 10^{-6} 
&  \frac{ 1}{96 \pi} y_c  y_u   K_{cb} y_{b}^2 K_{ub} 
  \sim   2\times
  10^{-17} t_\beta^2   
 \\
 & &\\
\hline
\tau \to \mu \gamma  & 10^{-4} 
&  \frac{ 1}{96 \pi}   y_c  y_t  K_{cb} y_{b}^2 K_{tb}  
   \sim  7 \times  10^{-10} t_\beta^2   
 \\
 & &\\
\hline
\end{array}
$
\caption{ The largest predicted  coefficient $\varepsilon$,  induced by
any of the  scalar leptoquarks we consider, of mass $\sim$ 300 GeV,
with the flavoured coupling $\lambda^{LQ}$ following
the patterns considered in section \ref{unit}. 
The second colomn is the bound on $\varepsilon$
(defined in eq. (\ref{defve})) for the process in the first
colomn.
The bounds above the double line are on pseudoscalar operators
(which can be induced by the two interactions of $S_0$ and $S_2$),
those below are $V\pm A$.
In the third colomn, $\lambda_{S_0}, \lambda_{S_2}  \lsim 0.25$. 
The expectation quoted for $\mu -e$ conversion is from a loop
diagram; the tree level expectation  is smaller, as discussed
around eq. (\ref{loopmec}). 
}
\label{results3}
\end{center}
\end{table}

\subsection{Some comments}
\label{commentaires}

In our estimates, and in the table \ref{results3}, we quote
the largest rate mediated by any of the leptoquarks. This is
not because all the leptoquarks are present with $m_S \sim 300$ GeV,
but rather that we prefer to present one concise table, 
rather than  one for each
of the six leptoquark couplings. 
In this
section, we assume the presence at the ``flavour scale''
of the new flavour structures (``spurions'') corresponding
to the identity matrix linking the $u$-type and lepton flavour spaces. 
With these new spurions, we can construct $\lambda$ matrices
for any leptoquark and guestimate the induced rates. 
The largest rates, for any leptoquark, are in the table. 
So if  only one, or some of the leptoquarks are
light, not all our guestimates will be fulfilled. 
For example, the bound on $\tan \beta$ from eq.
(\ref{Bnue}) only applies if there is an $S_0$ with
both chiral couplings and $m_S \sim 300$ GeV.

Rather than assuming that the new spurion was the identity
matrix, with a coefficient $\lambda_{S_{0, 2}}$ that we are free to ajust, 
 we  could take the approach that the spurions
were $\lambda_{S_0} {\cal I}$ and  $\lambda_{S_2} {\cal I}$.
Then the  operator coefficients  in eq. (\ref{opS0}) to eq.  (\ref{optS2}) would
all be proportional to the same coefficient $\lambda_{S_{0,2}}^2/m_S^2$. 
In table \ref{results3},  this would  multiply the 
coefficients of  pseudoscalar operators (above the double bar)
 by $\lambda_{S_{0,2}} \sim 1/4$, 
and the coefficients of  vector operators (below the double bar) 
would be multiplied by $\lambda_{S_{0,2}}^2 \sim 1/16$.
The ratio  eq. (\ref{Bnue}) 
is suppressed by a further factor of order of 
$\lambda_{S_0}^2\leq 1/16$, making the upper bound 
on $Br(B^+\to e^+\nu)$ compatible with $\tan\beta=O(1)$. 
The contributions from 
the operators suppressed by four powers of Yukawa couplings, 
already negligible, are even more suppressed in this case. 
The most sensitive probe would be $K^+\to\pi^+\nu_\tau\bar\nu_\tau$, 
in agreement with the current bound for  all values of $\tan\beta$.

An interesting question which models can address,
is the relative sensitivity of $B$ and $K$ decays to leptoquarks. 
The experimental bounds on the various $\varepsilon$s
arising from $K$ decays are lower than those from { $B$ mesons}.
However, leptoquarks could have larger couplings to third generation
fermions, resulting in  larger contributions in $B$ decays
than $K$ decays.  As discussed  around eq. (\ref{Bnue}), the ${\cal O}(y_f^2)$
charged current pseudoscalar decays, which can be $\propto y_\tau K_{tx}$,
are more tightly constrained by $B$ decays than by $K$ { mesons, whereas}
the best bound on the 
 ${\cal O}(y_f^4)$ operators, which can be $\propto y_\tau^2 K_{tx} K_{ty}$,
is from $K \to \pi \nu \bar \nu$. This illustrates the
interest of this decay for Beyond the Standard Model physics: 
it can probe the interactions of  third generation leptons, and
also of the top via the CKM matrix.

\begin{figure}[t]
\begin{center}
\epsfig{file=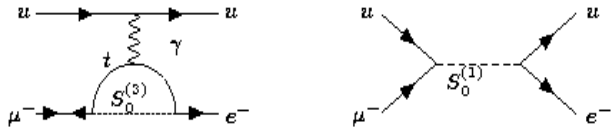, height=3cm,width=10cm}
\end{center}
\caption{Loop and tree leptoquark contributions to $\mec$. The
diagram represents leptoquarks with  quark flavour (= generation number), 
as  considered in section \ref{qflav} . 
For the various
patterns of $\lambda$'s that we consider, the loop diagram involving
third generation quarks dominates.
\label{fig1}  }
\end{figure}

The final issue is the relative importance
of loop and tree diagrams. With the hierarchical
couplings we consider, it is possible that (third
generation) loops dominate over (first generation)
tree level processes. For instance,  this occurs in  
$\mec$, which is well known to be a sensitive
probe of the effective  $\mu$ - $e$ -$\gamma$ vertex.
The  tree level amplitudes for $\mec$  give
\beq
6 \epsilon \simeq 
\left\{ \begin{array}{lclc}
 y_u y_d y_c K_{cd} y_d &\sim & 2\times 10^{-18} t_\beta^2& ~~~(\tilde{S}_2, \tilde{S}_0) \\
  y_e y_u y_\mu y_c K_{cd}  &\sim &   10^{-18} t_\beta^2& ~~~({S}_2)   
\end{array} \right.
\eeq
and are  smaller than  loop leptoquark
exchange (see the diagrams of figure \ref{fig1}).
We take the loop contribution to be \cite{ybook} 
 $|\varepsilon^{e \mu}|^2$  $ \times \alpha_{em} \log(m_W^2/m_\mu^2)$,
and obtain
the  loop to tree  ratio (mediated by $\tilde{S}_0$): 
\beq
\alpha_{em} \log \left( \frac{m_W}{m_\mu} \right)^2
\left| \frac{ \frac{N_c}{96 \pi^2}
 y_c  y_u  K_{cb} y_{b}^2 K_{u b}}
{y_u y_d y_c  y_d  K_{cd} } 
 \right|^2 
\sim  10^{-3}
\left| \frac
{  K_{cb} y_{b}^2 K_{u b}}
{  y_d^2 K_{cd} }  \right|^2 
\gsim 10 ~~.
\label{loopmec}
\eeq
In table \ref{results3}, we quote the loop expectation.

It can also arise in the SM that third generation loops
exceed tree level amplitudes (for instance in
$\bar p  p \to$ Higgs). This problem is exacerbated
for our $\lambda$ couplings, because they  can be proportional
to (Yukawa coupling$)^n$, for $n \geq 2$. This raises two questions.
First, have we considered the most restrictive loop-induced processes,
and secondly, are there loop contributions which exceed
the tree estimates in the tables. With respect to the first
question, we expect that
the  most sensitive one  loop diagrams
will have external lepton legs and an internal leptoquark
and $t$ or $b$ loop, because the coloured loop is enhanced by
$N_c$, and there are several strict limits on
New Physics  in the lepton sector. So  $\tmg$, $\meg$ and $g-2$
should give the best limits. We did not consider
the box contributions to meson-anti-meson mixing, 
which would be proportional to Yukawa eigenvalues for
the external quarks and for the  internal (third generation)
leptons.

It is difficult to ascertain whether there could
be loop diagrams larger than our tree estimates. 
To avoid the suppression present in
the tree amplitudes, due to small Yukawa eigenvalues,
the leptoquark  should not interact with the external
leg fermions. This would not be possible for
the lepton flavour violating decays, because 
the lepton flavour violation is provided
by the leptoquarks.  In table
\ref{results3}, that leaves
$K^+ \to \pi^+ \nu \bar\nu$ and
$B_s \to \mu^+ \mu^-$. 
An  SM loop would be required to induce the 
flavour off-diagonal quark current, so
the leptoquark loop would merely
modify flavour diagonal lepton
interactions already present in
the SM.  Such leptoquark
loops  should be
better constrained by  precision
observables, such
as $g-2$,  which probe lepton
interactions more directly.
So even if there is a loop
contribution that exceeds our tree
estimates, we doubt that
it would be  phenomenologically relevant.

\section{Leptoquarks with quark flavour}
\label{qflav}

In R-parity violating supersymmetric (RPV SUSY) theories, the squarks 
can have interactions with quarks and leptons. Insofar as a leptoquark
is a boson interacting with a lepton and a quark, such
squarks can therefore be identified as leptoquarks
(with a quark flavour index). 
An implementation of the MFV hypothesis  in RPV
SUSY seesaw
models has recently been investigated  by 
  Nikolidakis and Smith  \cite{NS}, who
showed that the lepton number violating couplings were sufficiently
suppressed that $R$-parity was not required to ensure
proton stability.

The idea  used by  Nikolidakis and Smith  to obtain spurions with a single
lepton flavour index, was the cross product.
Since SM flavour spaces are 3 dimensional,
the fully anti-symmetric 
$\varepsilon_{IJK}$ tensor 
can be contracted with an anti-symmetric  two-index 
object (such as $Y_e Y_e^\dagger [m_\nu]$, where 
$[m_\nu]$ is the symmetric  majorana mass
matrix of the light neutrinos) to obtain a spurion
with a singlet lepton index
\beq
\bUI \equiv \frac{1}{m_{atm}}
\varepsilon_{IJK} [ Y_e Y_e^\dagger m_\nu ]^{JK}
\label{defbU}
\eeq
where $m_{atm} = \sqrt{ \Delta m^2_{atm} } \simeq .05$ eV
is the atmospheric mass difference which we take as
the neutrino mass scale.

This formula requires some discussion, because  it is
phenomenologically 
peculiar to promote  the coefficient of the  non-renormalisable operator
$[m_\nu]$ to the status of fundamental
flavour structure (or spurion):  non-renormalisable operators
are not spurions 
in the quark sector, where   MFV is approximately
confirmed by the data. It is also theoretically
peculiar:  one can anticipate
that the flavour pattern was generated at some high scale,
and   transmited to low energy via renormalisable
couplings. So the flavour pattern in $[m_\nu]$ can arise
from the product of several spurions (as in the seesaw mechanism),
and the overall magnitude  is controlled by a ratio
of energy scales, which may have nothing to do
with the flavour structure.

We will use $[m_\nu]$ as a spurion anyway, because
a product of spurions is also a spurion, the neutrino
mass matrix is the only available information
about lepton flavour violation, and an object
with two indices in lepton
doublet space is required to contract with
$\varepsilon_{IJK}$.  However,
we normalise to the light neutrino mass scale $m_{atm}$,
rather than to the Higgs vacuum expectation value
 as in \cite{NS}, because
we think MFV is about the  flavour pattern, whereas
majorana neutrino masses may be small because they
violate lepton number\footnote{With our normalisation,
the demonstration of \cite{NS} that MFV suppresses RPV
sufficiently  would no longer hold.}. Phrased another way: the
ratio $\langle H \rangle/M_{\nu_R}$ in the seesaw mechanism
may have  nothing to do with flavour.

For hierarchical \footnote{For the inverse
hierarchy, we would obtain $\bU \sim
y_\tau^2 ( 1, \delta^2, \delta^2 y_\mu^2/y_\tau^2)$.}
  $\mnu$, and tri-bi-maximal mixing \cite{PDB}, this gives
\beq
\bar{\Upsilon}^I = 
\frac{\varepsilon^{IJK} [Y_e Y_e^\dagger m_\nu]_{JK}}{m_{atm}} =
\frac{1}{8}
\left[
\begin{array}{c}
  (y_\mu^2 - y_\tau^2) (4  -3  \delta)  \\
-\sqrt{2} (y_\tau^2 - y_e^2) (\sqrt{3} \delta - 4s_{13}) \\
- \sqrt{2} (y_\mu^2 -y_e^2)   (\sqrt{3} \delta + 4s_{13}) 
\end{array}
\right]
\sim - \frac{1}{2}
\left[
\begin{array}{c}
 y_\tau^2 \\
 y_\tau^2 \delta   \\
 y_\mu^2\delta
\end{array}
\right]
\label{eq20}
\eeq
for $\delta \equiv m_{sol}/m_{atm} \simeq 1/6$ and 
$s_{13} = \sin \theta_{13} \lsim .1$.
We can already anticipate that in this pattern,
a leptoquark interacts with all flavours of leptons,
and  most strongly to the first and second generation
doublets (there will be an additional $Y_e$
in the couplings to singlet charged leptons).
This  is unsurprising, since the lepton flavour violation
is related to the large mixing angles of the
lepton mixing matrix.  However,   lepton
flavour changing rates are suppressed by at least
a factor $y_\tau^4$.


The  spurion $\bUI$, of the authors of
\cite{NS}, which has only one
lepton doublet index, allowed them to construct
the $R$-parity violating $\lambda'LQD^c$ as
\beq
\lambda' \propto \bU Y_d
\eeq
In an analogous fashion, using $\bUI$,
 we can 
construct  the $\lambda$ couplings of leptoquarks
carrying quark generation number.
To do this, we must 
assign the leptoquark to live in
one of the quark flavour spaces. 
 A  possibility
would be for the leptoquark to carry the flavour  of
the quark with which  it interacts,  in which
case  no quark Yukawa matrix is required at the vertex
(the identity matrix is sufficient). 
Since we explicitly wish the leptoquark couplings
to have  flavour structure, we do not consider this option ---
although it is interesting because it could
give large couplings.
We discuss two possibilities  below, where the
leptoquark coupling matrices are proportional to quark Yukawa matrices. 

\subsection{Maximal coupling}
\label{max}
The largest leptoquark couplings are obtained
by allowing only one power of the quark  Yukawa coupling at the
vertex, selected according to the type of quark.
Suppose, for instance, that the leptoquark
interacts with singlet $d$ quarks. 
Then  $Y_d$ imposes that  $S$ has
doublet quark flavour indices.  With this
``maximum coupling'' hypothesis, we obtain
\bea 
\label{mcfirst}
{  {\bf \lambda_{L S_0}}} S_0  \overline{\ell} i \tau_2 q^c
& \to & 
 {\bf [ \bU^* ]}^I  {\bf [Y^\dagger_{d,u}]}^{rP} S_0^{ r}    \overline{\ell}_I i \tau_2 q_P^c
~~~~~~~~ S_0 \sim d, u
\\
{ {\bf \lambda_{R S_0}}} S_0  \overline{e} u^c 
 & \to & 
 {\bf [Y_e^T \bU^* ]}^{i}  {\bf [Y_u]}^{Pq} 
S_0^{ P}    \overline{e}_{i} u^c_q 
~~~~~~~ S_0  \sim q
\\
{  {\bf \tilde{\lambda}_{R \tilde{S}_0}} } \tilde{S}_0 \overline{e} d^c  
 & \to & 
  {\bf [Y_e^T \bU^* ]^{i} [Y_d]}^{Pr}    
\tilde{S}_0^{ P}  \overline{e}_{i} d^c_r
~~~~~~~ \tilde{S}_0  \sim q
\\
{ {\bf \lambda_{L S_2}} }  S_{2} \overline{\ell} u
 & \to & 
 {\bf  [ \bU^* ]}^I  {\bf [Y^*_u]}^{Pr}     S_{2}^{ P} \overline{\ell}_I u_r
~~~~~~~~~S_2 \sim \bar{q}
\\
 { {\bf \lambda_{R S_2}} } S_{2} \overline{e} [ i \tau _2 q ]  
 & \to & 
  {\bf [Y_e^T \bU^* ]}^{i} {\bf [Y^T_{u,d} ]}^{rP}   
S_{2}^{ r}  \overline{e}_i [ i \tau _2 q_P ]  
~~~~~~~~S_2 \sim  \bar{u}~ {\rm  or} ~ \bar{d}
 \\
{ {\bf \tilde{\lambda}_{L \tilde{S}_2}} } \tilde{S}_2
  \overline{\ell} d 
 & \to &
 {\bf [ \bU^* ]}^I  {\bf[Y^*_d]}^{Pr}  \tilde{S}_2^{P}  \overline{\ell}_I d_r 
~~~~~~~~~\tilde{S}_2 \sim \bar{q}
\label{mclast}
\eea
where we added the appropriate quark flavour
indices  to the  leptoquarks, and indicate in the
last colomn the flavour space they live in.
The  interactions involving the doublet quarks
${\bf \lambda_{LS_0} , \lambda_{R S_2}}$ can
be taken $\propto {\bf Y_d}$ or ${\bf Y_u}$; we take  ${\bf Y_u}$ because
the eigenvalues are larger.  The leptoquarks 
 $S_0$ and $S_2$  
both can  have two distinct interactions.
In this ``maximal'' pattern, the two interactions 
assign the leptoquarks to
different flavour spaces, which disfavours
this pattern, or the  presence of both couplings. 

This naive attempt to obtain large couplings leads to 
a peculiar behaviour:
the $U_Y(1)$ charges of the leptoquarks do not 
match those of the corresponding quarks with the same  
flavour structure. This is not an internal inconsistency:
the original $U(3)^5$ symmetry can be decomposed into five distinct
$U(1)$ and $SU(3)$ subgroups~\cite{dAGIS}, and we are not 
forced to choose the same $SU(3)$  and $U(1)$ assignments
for quarks and leptoquarks. 
However, it is clearly an unusual choice.
A flavour assignment for the leptoquarks that match 
their gauge quantum numbers will be discussed in the 
next subsection.

In this pattern, all generations of leptoquarks
 have 
similiar  couplings ($\lsim y_\tau^2 y_t$)
to all flavours of leptons. However, the couplings
to quarks are hierarchical and generation
diagonal (between the quark and
leptoquark), up to insertions of CKM matrix
elements. For instance,  third generation
leptoquarks  only interact with
 third generation
quarks,  and  doublet down-type quarks of the
first and second generations via CKM-suppressed terms.

The absence of a spurion which links
quarks and leptons has two consequences:
fermion generation diagonal interactions
are not favoured, and    the Yukawa suppression
of  quark and lepton bilinears can be  studied separately.
The $\varepsilon$ factor  (as defined  in eq. (\ref{defve}))
for a dimension six operator formed  
from the product of two bilinears,  will be
the product of  the coefficients given for the two bilinears.
The quark bilinears induced by
leptoquarks of generation $T$ (or $t$), with their coefficients, are:
\beq
\begin{array}{cc}
 K^*_{TP} y^2_{u,T} K_{TS} (\overline{d}_P \gamma^\mu P_L d_S)   ~~,~~ 
K^*_{TS} y^2_{u,T}  (\overline{d}_S\gamma^\mu P_L u_T)  ~~,~~ 
 y^2_{u,T} (\overline{u}_T \gamma^\mu P_L u_T) ~~,~~
  y^2_{u,t}  (\overline{u}_t\gamma^\mu P_R u_t) 
&
S_0 \\
 K^*_{TP} y^2_{u,T} K_{TS} (\overline{d}_P \gamma^\mu P_L d_S)   ~~,~~ 
 y^2_{u,T} (\overline{u}_T \gamma^\mu P_L u_T) ~~,~~
  y^2_{u,t}  (\overline{u}_t\gamma^\mu P_R u_t) 
& 
S_2 \\
 y^2_{d,t} (\overline{d}_t \gamma^\mu P_R d_t)  ~~,~~
& 
\tilde{S}_2 ~,~ \tilde{S}_0
\end{array}
\label{bilins}
\eeq
where $s,t$ ($T,P,S$) are  singlet (doublet) generation
labels, not to be summed over. As expected,  since we construct the quark
flavour structure of the leptoquark coupling with SM Yukawa matrices,
we find an MFV-like  suppression: quark flavour change can only
occur in charged current interactions, or among
the $d_L$s.
Since the two possible interactions
 of  $S_0$ and $S_2$ assign the leptoquark to different
flavour spaces, they cannot be simultaneous present,
and   bilinears like $ \overline{q}d$,
or $\bar{\ell}\sigma e F$   cannot
be generated (except with an external mass insertion). 

The lepton bilinears, with the flavour factors
of their  coefficients, are
\bea
\bU^{I*} \bU^J (\overline{\ell}_I \gamma^\mu \ell_J) ~~~,~~~
y_{e,i} y_{e,j}{\bU^{i*}} \bU^j (\overline{e}_i \gamma^\mu e_j)
\label{bilinL}
\eea
where $I,J,i,j$ are not summed.  Lepton flavour violation
is suppressed by an extra lepton Yukawa coupling in the singlets,
so focussing on the doublet bilinears which
are all suppressed by an overall $y_\tau^4/4$,  
the relative ratio of flavours is
\footnote{For the inverse hierarchy, we would
obtain
$ ee:e\mu: e\tau : \mu \mu : \mu \tau  \sim $
$1 : \delta^2 :   \delta^2 \left( \frac{m_\mu}{ m_\tau}
\right)^2 :\delta^4:\delta^4 \left( \frac{m_\mu}{ m_\tau} \right)^2 
$, which is sufficiently similar that we do
not consider it further.}
\bea
ee:e\mu: e\tau : \mu \mu : \mu \tau = 
1 : \delta :  \left( \frac{m_\mu}{ m_\tau}
\right)^2  \delta:\delta^2:\delta^2 \left( \frac{m_\mu}{ m_\tau} \right)^2 
~~~~~~~~~~~~{\rm normal~hierarchy}
\label{nh}
\eea 
where $\delta \simeq 1/6$ is defined after eq.  (\ref{eq20}).
This pattern predicts $BR(\meg) > BR(\tmg)/BR(\tau \to \mu \nu \bar\nu)$,
and for  a top and third generation ($S_2$) leptoquark
in the loop,
\beq
\varepsilon_{L,R}^{e \mu} \sim  \frac{1}{4} \frac{N_c}{64 \pi^2} y_\tau^4 y_t^2 \delta
\simeq 2 \times 10^{-4} \frac{m_\tau^4}{v^4} \tan^4 \beta  ~~.
\label{epsmegmax}
\eeq
For $S_0$ in the loop, we estimate \cite{Lavoura} $\varepsilon_{L,R}^{e \mu}$
a factor $\sim 1/3$ smaller.
This is within the current experimental bound $\varepsilon \lsim 10^{-6}$
for $\tan \beta \lsim 25$. 
This  dipole induced by
the third generation leptoquark,  will give the dominant
contribution to $\mec $ (see eq. \ref{loopmec})
because  the tree level  exchange of a first generation
leptoquark  is
suppressed by first generation quark Yukawa couplings (we estimate
the tree-level $\varepsilon \sim y_\tau^4 \delta y_d^2/24$).

In  table \ref{results4}, 
various rare decays are estimated. 
Generation number is not conserved in this pattern, so the
experimentally prefered rare decay
 $B_s \to \mu \bar{\mu}$ is also the more sensitive one
\beq
\frac{BR(B_s \to \mu^\pm \tau^\mp)}
{BR(B_s \to \mu^\pm \mu^\mp)} \sim 
\frac{| y_\mu y_\tau y_\tau^2 \, y_\mu^2 \, \delta^2  ~ K_{ts} K_{tb}|^2} 
{ |y_\mu^2 y_\tau^4 \, \delta^2  ~ K_{ts} K_{tb}|^2} \ll1
\label{Btmu/Bmumu}
\eeq
However, the meson decay rates in this pattern are
very small. From eq. (\ref{bilins}), one sees  that
no flavour change in induced at tree level among  singlet
quarks, or up-type doublet components. 
The 
FCNC decays  of $K$ and $B$ mesons only occur through
the first  bilinear  of eq. 
(\ref{bilins}), which combines 
with $(\bar{\nu} \gamma_\mu P_L \nu)$ or
 $(\bar{e} \gamma_\mu P_R e)$
in the four fermion interactions induced at
tree level by
$S_0$s and $S_2$s (see
table \ref{4fv}).  Hence in
eq. (\ref{Btmu/Bmumu}), the charged leptons are
singlets, and have the additional $y_{e,i} y_{e,j}$ 
factor of the second bilinear of eq. (\ref{bilinL}).
As can be seen from table \ref{results4},
the least suppressed  meson decay is  $K^+ \to \pi^+ \nu \bar\nu$,
mediated by $S_0$, 
which reaches the experimental bound for $\tan \beta \sim 100$. 
However, as shown in eq.  (\ref{epsmegmax}),
 $\meg$, which can also be  mediated by $S_0$, is detectable for
 $\tan \beta \sim 35$.  

The most promising precision  searches
for this pattern of leptoquark couplings would be $\meg$ (or
$\mec$).  

At hadron colliders, leptoquarks can be produced
via strong interactions, and  decay via their $\lambda$
couplings.  Searches frequently suppose that  
$\lambda   \gsim 10^{-8}$,  so that  the  leptoquark
decays within a few centimetres of its production point. 
This condition is verified, for $\tan \beta \geq 1$,  for all the
third generation leptoquarks  in this pattern
except  $\tilde{S}_0$ (which requires  $\tan \beta \gsim 2$).
So searches for leptoquarks decaying to
a $t$   or $b$ are  interesting.
The third generation quark would be accompagnied by
an electron or a muon, due to the comparatively
democratic coupling to leptons in this pattern, 
as  can be
seen from equations  (\ref{mcfirst}) to (\ref{mclast}).

\renewcommand{\arraystretch}{1.45}

\begin{table}
\begin{center}
$
\begin{array}{||l|l|l|l||}\hline
{\rm process} & \varepsilon < & {\rm S~ with~ quark~ flav~ (max)} & {\rm 
S~with~quark~ flav ~(gauge)} \\ 
\hline \hline
D^+ \to \pi^+ \mu^\pm e^\mp  & 2 \times 10^{-2}  
&& \frac{1}{24} K_{cb}y_b^2 K_{ub} y_\tau^4 \delta  \sim  7 \times  10^{-18} t_\beta^6  
 \\ 
 &  & & \\
\hline
%
%
K^+ \to \pi^+ \nu \bar{\nu} & 9 \times 10^{-6}  
& \frac{1}{24}y_\tau^4 y_t^2 K_{ts} K_{td}  \sim  10^{-13} t_\beta^4  
& 
\\
   & & &\\
\hline
K_L \to \mu^\pm e^\mp &3 \times 10^{-7} 
&  \frac{1}{24}y_\tau^4 y_\mu y_e \delta K_{ts} y_t^2 K_{td} 
  \sim  5 \times 10^{-23} t_\beta^6  
&
\\
   & & &\\
\hline
B_s \to \mu^\pm \mu^\mp & 7 \times 10^{-5} 
&  \frac{1}{24} y_\tau^4 y_\mu^2 \delta^2 K_{ts} y_t^2 K_{tb}
 \sim  10^{-19} t_\beta^6   
&
\\
   & && \\
\hline
B^+ \to K^+ \tau^\pm \mu^\mp & 2 \times 10^{-3}  
&  \frac{1}{24} y_\tau^3y_\mu^3 \delta^2 K_{ts} y_t^2 K_{tb} 
 \sim  10^{-20} t_\beta^6  
&
\\
   & & &\\
\hline
\mu N \to e N' &8 \times 10^{-7} 
& 
 \frac{1}{256 \pi^2} y_\tau^4 y_t^2 \delta
\simeq   10^{-12}  t^4_\beta &
 \frac{1  }{16 \pi^2} t_\beta y_\tau^4 \delta  y_b^2
  \sim 8 \times 10^{-15} t_\beta^7
\\
   & & &\\ 
\hline
\mu \to e \gamma  &  \sim 10^{-6} 
&   \frac{1}{256 \pi} y_\tau^4 \delta  \sim 2 \times  10^{-12} t_\beta^4  
&  \frac{1  }{16 \pi} t_\beta y_\tau^4 \delta  y_b^2
  \sim 2 \times 10^{-14} t_\beta^7  
 \\
 & & &\\
\hline
\tau \to \mu \gamma  & \sim 10^{-4} 
&  \frac{1}{256 \pi} y_\tau^2 y_\mu^2 \delta^2 
   \sim  10^{-15} t_\beta^4  
&  \frac{1 }{16 \pi} t_\beta 
y_\tau^2 y_\mu^2 \delta^2  y_b^2 
  \sim  8 \times 10^{-18} t_\beta^7  
\\
 & & &\\
\hline
\end{array}
$
\caption{ Predicted   coefficients $\varepsilon$, induced by
a scalar leptoquark of mass $\sim$ 300 GeV,
with the flavoured coupling $\lambda^{LQ}$ 
arising when leptoquarks carry quark flavour.  
The second colomn is the bound on $\varepsilon$
(defined in eq. (\ref{defve})) for the process in the first
colomn, and the third and fourth  colomns are 
the largest expected values of $\varepsilon$
mediated by any  of the leptoquarks, for
respectively the  cases considered in
sections \ref{max} and \ref{gauge}. 
  }
\label{results4}
\end{center}
\end{table}

\subsection{Coherent gauge and flavour assignments}
\label{gauge}

In this section, we suppose that the 
flavour space of the leptoquarks is  determined
by their gauge couplings. For instance,  doublet
leptoquarks should be in the $q$ flavour space,
and  the hypercharge of $S_0$  
implies that it should  live
in $d$  space.
This suggests  the following leptoquark interactions:
\bea 
  {\bf \lambda_{L S_0}}  S_0 \overline{\ell} i \tau_2 q^c
& \to & 
 {\bf [ \bU ^*]}^I{\bf [Y^\dagger_{d}]}^{rP} 
S_0^{ r}  \overline{\ell}_I i \tau_2 q^c_P
~~~~~~S_0 \sim  {d}
\label{s1}
\\
 {\bf \lambda_{R S_0}} S_0   \overline{e} u^c 
 & \to & 
   {\bf [ Y_e^T \bU^*]}^{i} {\bf  [Y_{d}^\dagger Y_{u}]}^{pr} 
S_0^{ p}  \overline{e}_i u^c_r 
~~~~~~S_0 \sim  {d}
\label{s2}
\\
 {\bf \tilde{\lambda}_{L \tilde{S}_2}} \tilde{S}_2   \overline{\ell} d  
 & \to &
 {\bf [ \bU^* ]}^I{\bf [Y^*_d]}^{Pr}  
\tilde{S}_2^{ P} \overline{\ell}_I d_r 
 ~~~~~~~~~~\tilde{S}_2 \sim \bar{q}
\eea
The hypercharge of $S_2$ and $\tilde{S}_0$  do not match that
of any SM coloured particles, so we do not consider them further. 
So $S_0$ and $\tilde{S}_2$  ressemble, respectively, 
a singlet $d$ squark,  and a anti-squark
doublet, and the interactions of these ``leptoquarks''
should correspond to  those of  squarks with the
R-parity violating superpotential
$ W  \supset \lambda'_L LQD^c + \lambda'_R D^{c*} E^c U^c$
(where we allow the non-holomorphic interaction corresponding
to $\lambda_{R S_0}$).

The coefficients of the lepton bilinears of eq. (\ref{bilinL})
remain the same. The chirality flipping dipole operator
$ \overline{\ell}_J \sigma^{\a \b} e_i F_{\a \b}$
can now arise,  due to the simultaneous presence
of $\lambda_{L S_0}$, and $\lambda_{R S_0}$.
 From eq. (\ref{f1f2g}), with a top and third
generation leptoquark in the loop:
\beq
\varepsilon 
\sim  4  \frac{m_t}{m_{e,i}} \frac{N_c}{16 \pi^2}
\bU^J \bU^i y_{e,i}\,   y_b^2\,  
\to  \tan \beta \,  \left[ y_b^2\,   \frac{N_c}{16 \pi^2} y_\tau^4\,   \delta
\right]
\eeq
where after the arrow is the expectation for $\meg$.
This is enhanced by a single
$\tan \beta$  with respect to the  $\varepsilon$ obtained 
with  an
external  $m_{\mu}$ insertion, and has
a larger numerical factor.
For $\tan \beta \sim 1$, 
this  is  less than the  expectation 
in the previous ``maximal'' pattern
(see eq. (\ref{epsmegmax})),
due to the additional factors of $Y_d$. 
However,  $\varepsilon \propto \tan^7 \beta$
grows rapidly with $\tan \beta$, and would  exceed
the current bound on $\meg$  for $\tan \beta \gsim 10$.

Unlike the ``maximal'' pattern of section \ref{max},
the coefficients of quark bilinears are flavour changing
for doublet and singlet up-type quarks, as well
as the charged current. 
 Since the $S_0$  and $\tilde{S}_2$ leptoquarks
carry  ``down-type'' flavour, they do not mediate
flavour-changing interactions among down-type quarks
at tree level
\footnote{If  the possibility of 
 generation-mixing via the leptoquark
mass matrix was included --- as happens for instance
for squarks---this would no longer be the case.}.  
The induced quark bilinears are for $S_0$ are
\beq
 [K  D_d^2 ]_{PS} (\overline{u}_P \gamma^\mu P_L d_S)  ~~  , ~~
 [K  D_d^2 K^\dagger]_{PS} (\overline{u}_P \gamma^\mu P_L u_S)  ~~  , ~~
 [D_u K D_d^2 K^\dagger D_u]^{ps}  (\overline{u}_p \gamma^\mu P_R u_s) ~~,~~
 y^2_{d,S} (\overline{d}_S \gamma^\mu P_L d_S) ~~(S_0),~~
\label{e46}
\eeq
and for $\tilde{S}_2$:
\beq
 y^2_{d,s}
(\overline{d}_s \gamma^\mu P_R d_s)  ~~(\tilde{S}_2),~~
\label{e47}
\eeq
where $p,s$ ($P,S$) are  singlet (doublet) generation
labels, not to be summed over.
Flavour changing  $V \pm A$ bilinears among the  up-type quarks,
 are  suppressed
by $y_{f}^2$ for doublets and $y_f^4$ for singlets, so
neutral pseudoscalar meson decays induced in
this pattern with be undetectable: $B$ and $K$ decays
do not arise, and $D \to \mu \bar{e}$ is small. 
Any quark
bilinear is suppressed by $y_f^2$ or $y_f^4$, as compared
to the (undetected) interactions  of the Higgs,
which couples to quark bilinears with a single power
of $y_f$.  

Pseudoscalar operators are generated by the two
chiral couplings  of $S_0$, with quark bilinears
and coefficients:   
\bea
 y^2_{d, P}
K_{rP}^* y_{u, r}(\overline{d}_P P_R u_r) ~~,~~
K_{sP} y^2_{d, s}
K_{rs}^* y_{u, r} (\overline{u}_P P_R u_r)~~~~
\eea
The  flavour diagonal pseudoscalars   are undetectable
compared to the  pseudoscalar couplings
of the Higgs $\propto y_f$.  From
eq. (\ref{s1}) and eq. (\ref{s2}),
we see that $ \lambda_{LS_0}
\lambda_{RS_0} \sim y_{e,i} y_{u,r}\lambda^2_{LS_0}$,
so the contribution of the pseudoscalar operator
to pseudoscalar meson decays is smaller,
by the factor $y_{u,i}$, than
that of the $V -A$ operator.
Therefore
  in table
\ref{results4}, we do not  estimate
rates for  pseudoscalar
meson decays in this pattern.


 Some
estimates for $\varepsilon$s can be found in table \ref{results4}.
Since this ansatz does not induce tree
level FCNC among down-type quarks (no $B_s \to \mu \bar\mu$,
$K_L \to \mu^\pm e^\mp$), the most sensitive
rare decay  is $\meg$.

 Leptoquark  decay to a 
charged lepton ($e, \mu$) and a $t$ or $b$
is an interesting search channel for  this pattern 
at hadron colliders (similarly
to   the  ``maximal'' pattern
of couplings discussed in section \ref{max}).
 For third generation leptoquarks, the couplings $\lambda_{LS_0},
\lambda_{L \tilde{S}_2}
 \gsim 10^{-8}$ for $\tan \beta \sim 1$, so the leptoquarks  would
decay within a few centimetres of the production point. 
However, a third generation $S_0$ with only
the  coupling $\lambda_{R S_0}$, could  appear
as a track in the detector, since the largest
$\lambda \lsim y_\mu y_\tau^2 \delta y_b \tan^4 \beta
\sim 10^{-10} \tan^4 \beta$.
Lower  generation leptoquarks  have very small $\lambda$s,
potentially allowing them to hadronise and escape the detector.
However, we imagine that  in a more realistic
model, there would be intergeneration mixing among
leptoquarks, which could allow faster decays.


\section{Leptoquarks with quark and lepton flavour}

\label{lqflav}

The final possibility that we consider is to
attribute both quark and lepton flavour to the leptoquarks. 
There are numerous possibilities. To avoid listing them all,
 we  require, as discussed prior to section \ref{max},
that the $\lambda$ matrices be proportional
to quark Yukawa matrices,  and that hypercharge
survive as  a global symmetry  in the presence of
the leptoquark couplings $\lambda$ (see section \ref{gauge}). 
In practice, the second condition  requires the sum
of the hypercharges of the flavour spaces in which the
leptoquark lives should be the hypercharge of the leptoquark.
This allows the following  $\lambda$s:
\bea 
\label{ql1}
S_0  {\bf \lambda_{L S_0}} \overline{\ell} i \tau_2 q^c
& \to & 
S_0^{ jq}   {\bf [Y_e^*]}^{ Ij} {\bf [Y^*_u]}^{Pq} 
\overline{\ell}_ I i\tau_2 q^c_P
\\
S_0  {\bf \lambda_{R S_0}} \overline{e} u^c 
 & \to & 
S_0^{ jq}    {\bf [\tWe]}^{ij} {\bf [\tWu]}^{pq}     \overline{e}_i u_{p}^c 
\\
\tilde{S}_0 {\bf \tilde{\lambda}_{R \tilde{S}_0}}  \overline{e} d^c  
 & \to & 
\tilde{S}_0^{ jq}   {\bf  [\tWe]}^{ij} {\bf [\tWd ]}^{pq}    
\overline{e}_i d^c_{p}
\\
 {\bf \lambda_{L S_2}} \overline{\ell}  u  S_{2}
 & \to & 
 {\bf  [ Y_e^* ]}^{Ij}  {\bf [Y^*_u]}^{Pq}   \overline{\ell}_I  u_q  S_{2}^{ jP}
\\
 {\bf \lambda_{R S_2}}\overline {e}  [ i \tau _2 {q}]^T S_{2} 
 & \to & 
 {\bf  [ \tWe]}^{ji} {\bf [W_{u}]}^{QP}  
\overline{e}_i [{q}_Q i \tau _2]^T  S_{2}^{ j P}
\label{qfquest}
 \\
 {\bf \tilde{\lambda}_{L \tilde{S}_2}}  \overline{\ell} d  \tilde{S}_2
 & \to &
 {\bf [ \We ]}^{JI} {\bf [\tWd]}^{qp}  \overline{\ell}_I d_q  \tilde{S}_2^{ Jp}
\label{ql5}
\eea
Giving lepton and quark flavour to the leptoquarks will
lead to a multiplicity of leptoquarks: each of the five possible
singlet and doublet leptoquarks will come in 3 colours and 9 flavours.
This   can be  consistent with precision electroweak data,
if these particles obtain mass other
than by interacting with the Higgs, 
as discussed at the end of section \ref{notn}.

The quark and lepton indices of the $\lambda$s are unrelated
in this pattern, so the quark and lepton bilinears, and their
coefficients, can be studied separately.  In the mass
eigenstate bases of the quarks,
the bilinears are suppressed by two or four  powers of
Yukawa eigenvalues, and FCNC arise via CKM. The $V \pm A$
bilinears mediated by the nine types of $S_0$ are
\beq
 [ D_u^2 K]_{PS} (\overline{u}_P \gamma^\mu P_L d_S)  ~~ , ~~
 [K^\dagger  D_u^2 K]_{PS} (\overline{d}_P \gamma^\mu P_L d_S)  ~~ , ~~
 y^2_{u,S} (\overline{u}_S \gamma^\mu P_L u_S) ~~,~~
 y_{u,p}^4 (\overline{u}_p \gamma^\mu P_R u_p)  ~~  , ~~
\eeq
where the $(\bar{d} \gamma P_L d)$ bilinear  combines with
neutrinos, and the others with two charged leptons or a charged current
as required (see table \ref{4fv}).
The doublet $S_2$ gives
\beq
 [K^\dagger D^4_{u} K]_{PT} (\overline{d}_P \gamma^\mu P_L d_T) ~~,~~
y_{u,p}^4  (\overline{u}_p \gamma^\mu P_L u_p) ~~ ,~~
y_{u,p}^2  (\overline{u}_p \gamma^\mu P_R u_p) ~~ ,~~
\eeq
and both $S_0$ and $S_2$ can mediate 
pseudoscalar operators,
with coefficients   
\bea
K_{sP}^*  y^3_{u, s}(\overline{d}_P P_R u_s) ~~,~~
 y^3_{u, r}(\overline{u}_r P_R u_r) ~~
\eea
Finally the leptoquarks interacting with singlet
$d$s ($\tilde{S}_2, \tilde{S}_0$) induce:
\beq
 y^4_{d,s}
(\overline{d}_s \gamma^\mu P_R d_s)  ~~ .~~
\eeq

Lepton flavour is conserved for the
$\lambda$s of eq. (\ref{ql1}) to  eq. (\ref{ql5}), because 
the leptonic part of  the $\lambda$s is constructed
only with  $Y_e$. 
At tree level, this pattern  therefore
generates
 $ V\pm A$
four fermion operators that arise in
the SM, with coefficients $\propto y_f^4, y_f^6$
or $y_f^8$. These  can be compared to four fermion
interactions induced by the Higgs boson, which
have a coefficient $\propto y_f^2$ and are unobserved. 
 The 
 most sensitive process would be $B_s \to \mu \bar{\mu}$
(induced by a $V \pm A$ operator, because
the leptoquarks of eq. (\ref{BRW}) do not generate
a pseudoscalar operator $(\bar{d} P_R d)(\bar{e} P_L e)$,
see table \ref{4fv}). However,the
 $B_s \to \mu \bar{\mu}$ amplitude is suppressed by
an extra $y_\mu^2$ with respect to the SM, so
large $\tan \beta$ would be required to detect it. 
The pseudoscalar
operators are also sufficiently suppressed.

Leptoquark-quark loops can induce flavour
diagonal lepton dipole operators, 
such as $(g-2)_\mu$ \cite{Cheung:2001ip}. However, it is easy to
see that the contribution to  $(g-2)_\mu$ 
is always negligeable. The  one loop SM electroweak
contribution  $\simeq G_Fm_\mu^2/(8 \pi^2)$ is of
order the experimental uncertainty, and we can
guesstimate that the leptoquark loops
$\lsim N_c \lambda^2 m_\mu^2/(8 \pi^2 m_S^2)$.
Since the $\lambda$s which couple to muons
are proportional to $y_\mu$, this is very  small. 

Another potentially interesting  
process is $b \to s \gamma$;  
with a $\tau$ or $\nu_\tau$  in the loop.    
From eq.
(\ref{f1f2g}), we obtain  (for $S_0$ and $\nu_\tau$
in the loop)
\beq
\varepsilon \sim \frac{1}{6} \frac{1}{96 \pi^2}
y_\tau^2 y_t^2 K_{tb} K_{ts}
\eeq
Assuming that the leptoquarks
can contribute at most  $\sim 30 \%$ of the SM $b \to s \gamma$ rate,
we estimate $\varepsilon^{sb}  \lsim 2 \times 10^{-4}$.
As can be seen from table \ref{results5},
$b \to s \gamma$  is less sensitive than $B_s \to \mu \bar\mu$,
because the loop suppression more than compensates
for the larger $\tau $ Yukawas.

To obtain lepton flavour violation in this pattern, 
we can introduce the  lepton number conserving
spurion associated to the majorana mass matrix:
\beq
\widetilde{W}_\nu 
\equiv \frac{1}{m_{atm}^2}[m_\nu] [m_\nu]^\dagger
= U D_\nu^2 U^\dagger 
\eeq
where $U$ is the leptonic mixing matrix,
we assume a normal hierarchy for neutrino masses
so $D_\nu^2 = $ diag $\{ 0, \delta^2, 1 \}$,
and we  neglected the lightest neutrino mass. 
 
This additional spurion could multiply various
lepton doublet indices appearing in the construction of
the $\lambda$s. For instance, if we maintain
the lepton doublets in the phenomenologically
relevant charged lepton mass basis, we can
nonetheless perform the replacement
\footnote{ The replacement $\We \to  \widetilde{W}_\nu \We$ in  eq. 
(\ref{ql5}) is also possible, but for simplicity we
do not consider it.}:
\beq
\tWe \to Y_e^\dagger \widetilde{W}_\nu Y_e
\eeq
 in eq.
(\ref{ql1}) to eq. (\ref{ql5}) above,
which allows the  LFV bilinears
\bea
[D_e U D_\nu^2 U^\dagger D^2_e U D_\nu^2 U^\dagger D_e]^{ij}
(\bar{e}_i \gamma^\rho P_R e_j)
\simeq y_{e,i} U_{i3} \frac{y_\tau^2}{2} U_{j3} y_{e,j}
(\bar{e}_i \gamma^\rho P_R e_j)  \\
{[ D^2_e  U D_\nu^2 U^\dagger D_e ]^{Ij}}
(\bar{e}_I P_R e_j)
\simeq y^2_{e,I} U_{I3}  U_{j3} y_{e,j}
(\bar{e}_I  P_R e_j)
\eea
The last approximation assumes that $U_{e3} = \sin \theta_{13}\gg
\delta^2 \simeq .03$.
 This allows the $\varepsilon$s listed
in the last colomn of table \ref{results5},
where we approximate $U_{\mu 3} \simeq U_{\tau 3} \simeq 1$.

\renewcommand{\arraystretch}{1.45}

\begin{table}[htb]
\begin{center}
$
\begin{array}{||l|l|l|l||}\hline
{\rm process} & \varepsilon < 
& S {\rm  ~with ~Q~ and ~ L ~flav}  &  S {\rm  ~with ~Q~ and ~ L ~flav,~ and~ LFV}\\ 
\hline \hline
K^+ \to \pi^+ \nu \bar{\nu} & 9 \times 10^{-6}  
  &  \frac{1}{6} y_\tau^2 K_{ts} y_t^2 K_{td}  \sim 5 \times 10^{-9} t_\beta^2  
&  \frac{1}{6} y_\tau^2 K_{ts} y_t^2 K_{td} \sim 5 \times 10^{-9} t_\beta^2  \\
 & & &\\
\hline
K_L \to \mu^\pm e^\mp &3 \times 10^{-7} 
 & 
&  \frac{1}{12} y_\mu y_\tau^2 s_{13} y_e  y_t^4 K_{ts}  K_{td}  
  \sim  \frac{s_{13}}{.1} \,  5 \times  10^{-19}\, t_\beta^4  \\
 & & &\\
\hline
B_s \to \mu^\pm \mu^\mp & 7 \times 10^{-5} 
 &   \frac{1}{6} y_\mu^4 y_t^4 K_{ts}  K_{tb}   \sim  10^{-15} t_\beta^4  
&  \frac{1}{6} y_\mu^2 y_\tau^2 y_t^4 K_{ts}  K_{tb}  \sim 2 \times 10^{-13} t_\beta^4   \\
 & && \\
\hline
B^+ \to K^+ \tau^\pm \mu^\mp & 2 \times 10^{-3}  
 & &    \frac{1 }{12}y_\mu y_\tau^3 y_t^4 K_{ts}  K_{tb}  
\sim   10^{-12} t_\beta^4\\
 & & &\\
\hline
b \to s \gamma   & \sim 2 \times  10^{-4} 
&  \frac{1}{576 \pi^2}   y_\tau^2   y_t^2 K_{tb} K_{ts} 
 \sim 7 \times 10^{-10} t_\beta^2  
&  \frac{1}{576 \pi^2}   y_\tau^2   y_t^2 K_{tb} K_{ts} 
 \sim 7 \times 10^{-10} t_\beta^2   \\
 & & &\\
\hline 
\mu N \to e N' &8 \times 10^{-7} 
& & \frac{1}{64 \pi^2} y_\mu y_\tau^2 s_{13} y_e y_t^2 
    \sim    2  \frac{s_{13}}{.1} \,  10^{-17} t_\beta^4  \\
 & & &\\ 
\hline
\mu \to e \gamma  &  \sim 10^{-6} 
 & & \frac{1}{64 \pi}  y_\mu y_\tau^2 s_{13} y_e  y_t^2
   \sim  \frac{s_{13}}{.1} \,    10^{-16} t_\beta^4 
 \\
 & & &\\
\hline
\tau \to \mu \gamma  & \sim 10^{-4} 
&&  \frac{1}{64 \pi}  y_\mu y_\tau^3   y_t^2
 \sim 2 \times 10^{-12} t_\beta^4 \\
 & & &\\
\hline
\end{array}
$
\caption{ The largest expected  amplitudes, induced by
any  scalar leptoquark  of mass $\sim$ 300 GeV,
with the flavoured coupling $\lambda^{LQ}$
arising when the leptoquark carries quark and lepton flavour.
The second colomn is the bound on $\varepsilon$
(defined in eq. (\ref{defve})) for the process in the first
colomn, the following colomns are  the expectations
respectively  with and  without lepton flavour violation
via the neutrino mass matrix.
  }
\label{results5}
\end{center}
\end{table}

In this pattern  which allows for
lepton flavour violation, 
the most sensitive rare decay  
would be $\tmg$,  where the predicted amplitude
(due to $S_2$ exchange) 
 becomes of order of the current bound  for $\tan \beta \sim 80$.
As in previous sections, the loop contribution to $\mec$,
of a third generation $S_0$ (with a $b$ in the loop)
or $S_2$ (with $t$ or $b$ in the loop), which is
listed  in table \ref{results5},  dominates over 
the tree contribution of an $S_2$ to $\mec$:
\beq
\varepsilon^{e \mu dd}  \simeq
\frac{1}{6} y_\mu y_\tau^2 s_{13} y_e K_{td}^2y_t^4 
    \sim  2 \frac{s_{13}}{.1} \,  10^{-19} t_\beta^4 ~~
~~~~ ~~~~~~~~{\rm (tree } ~, ~ S_2 )
\eeq
The estimates in table \ref{results5} 
show that leptoquarks with quark and lepton
flavour   remains difficult to detect in rare
decays, even with  the addition of lepton flavour violation.
However, if  intergeneration mixing among the
leptoquarks was allowed, as could be expected
in a realistic model of leptoquark masses,
the rare decay rates could be enhanced.

Leptoquarks with such a pattern of couplings
could have interesting signatures at colliders. 
In the absence of intergeneration mixing, the 
lower generation leptoquarks  could hadronize
and travel in the detector before they decay. 
However the third generation leptoquarks
decay promptly to  $t$s or $b$s and $\tau$s, 
and in the case of 
$S_0$ and $S_2$ leptoquarks of second  lepton and
third quark generation, the  decay
to a $\mu$  and a third generation quark
also takes place within  a few { centimetres} of the production point
for $\tan \beta \sim 1$. 



\section{Summary and Discussion}
\label{disc}

Data from rare decay searches and collider experiments
implies that 
leptoquarks  with 
$m_S \lsim$ TeV
 should not have 
${\cal O} (1)$ couplings to leptons and quarks of
arbitrary flavour.  This can be quantified
as constraints on a dimensionless $\varepsilon$ 
coefficient of dimension six operators,
as defined in equations (\ref{defve})  and (\ref{dipop}). 
 Indeed, it is well
known that  New Physics at the electrweak scale
should have  its  flavoured interactions   patterned on those
of the Standard Model. For several New Physics scenarios,
such as Supersymmetry, this can be
elegantly obtained by imposing Minimal Flavour Violation (MFV).
However, since the Standard Model does not
provide a template interaction (which could
serve as a ``spurion'') linking
leptons and quarks, it is not
obvious how to apply the elegant formulation
of  MFV of d'Ambrosio {\it et al.} \cite{dAGIS}  to leptoquarks.
Phrased another way, the leptoquark coupling
matrix $\lambda$ has  one lepton  generation index and one
quark generation index; how can this be constructed
from  the SM mass matrices, which have two lepton, or
two quark indices?
In this paper, we  explore three 
ways to construct the leptoquark-quark-lepton
couplings $\lambda$  out of the observed
mass matrices.  
For simplicity, to reduce the permutations,
we consider  only electroweak singlet and doublet
scalar leptoquarks. 

\begin{enumerate}
 \item In section \ref{unit},
the leptoquarks have neither lepton nor quark
flavour, but a new ``spurion'', or flavour structure,
is introduced.  It is a unit matrix,
because this is the most minimal of structures, and
it connects the mass eigenstate basis of singlet
$u$-type quarks, to the mass eigenstate basis of
charged leptons. This ensures that the new
spurion does not introduce any new bases in
the vector spaces of flavour, and avoids the stringent
bounds from $B$ and $K$ decays.  The  leptoquark
couplings to other types of quark or lepton can be
obtained by multiplying the unit matrix by Yukawa
matrices, as given in eq. (\ref{e17}) and eq. (\ref{e18}).

This pattern of leptoquark couplings has the
interesting feature of favouring interactions
between leptons and quarks of the same generation.
This is an ``intuitive'' expectation for leptoquark
couplings, which could be interpreted as a leptoquark signature. It is 
to avoid the strict constraint from $K_L \to \mu^\pm e^\mp$
that  the unit matrix connects singlet $u$-type quarks to leptons.

The (tree level) four fermion operator coefficients generated
in this pattern are given in eq. (\ref{opS0}) to eq. (\ref{optS2}).
Generation non-diagonal quark-lepton couplings can
arise due to CKM, in the presence of $d$-type quarks.  
Since the unit matrix connects
singlet $u$s to the leptons of the same
generation,  the $\lambda$ couplings of 
leptons to $d$-type quarks are proportional to quark
Yukawa matrices. Leptoquarks therefore have
stronger couplings to $b$ quarks than $s$ quarks;
$B \to e \nu_\tau ,  \tau \nu_e$ are among the most
sensitive  low energy processes, followed by
$K^+ \to \pi^+ \nu_\tau \bar{\nu}_\tau$.
In  the table \ref{results3}  are listed 
the $\varepsilon$ factors  for
various  rare processes which could be sensitive
to this  pattern of
leptoquark couplings.

\item The second prospect, explored in section \ref{qflav},
is to attribute quark generation number to the leptoquark.
The $\lambda$ couplings thus have one quark flavour index
for the leptoquark and one for the quark, so
can be  proportional to a quark Yukawa matrix. 
The single lepton index of $\lambda$ can be 
obtained following an idea of Nikolidakis and
Smith \cite{NS},  discussed around
eq. (\ref{defbU}), which combines the antisymmetrric 
$\epsilon_{IJK}$ with the majorana neutrino mass matrix
$[m_\nu]_{IJ}$. 

In this approach, we must choose the quark flavour
space in which to place the leptoquark.
The first possibility  which we study, in section
\ref{max}, is to choose the largest  Yukawa matrix
interacting with the quark at the vertex, which fixes the
flavour space for the leptoquark. However, a
more { ``consistent''} approach, 
studied in section \ref{gauge},
might be to place the leptoquark in the  flavour
space of quarks who have the same hypercharge.

Leptoquark couplings $\lambda$ constructed according to
this pattern do not relate the quark to
lepton generation indices. The interaction with
quarks is proportional to (one or two powers of) quark
Yukawa matrices, so they 
are  hierarchical, with flavour changing neutral 
currents suppressed as in the usual Minimal Flavour Violation. 
This can be seen from the coefficients of  quark bilinears,
which contribute to the four fermion operators induced by the leptoquarks,
and which are given in eq. (\ref{bilins}), eq. (\ref{e46}) and eq. (\ref{e47}).
The two cases we consider  differ in that the tree level FCNC 
are among $d$-type quarks  in section \ref{max},
and  among $u$-type quarks  in section \ref{gauge}.
All three generations of leptoquark have
similar interactions to  $e$s and $\mu$s, with
some suppression to $\tau$s (see eq. (\ref{nh})),
due to the democratic structure of the majorana
neutrino mass matrix, which provides the flavour violation.
Due to the hierarchy in couplings to quarks,
the most sensitive decay for these patterns
would be $\meg$ with a $t$ or $b$ in the loop.
 In  the table  \ref{results4} 
 the $\varepsilon$ factors for various 
other processes are listed.

\item   Finally in section \ref{lqflav} we consider
 leptoquarks carying lepton and quark generation
indices. This implies a large number of leptoquarks
(3 colours $\times$ $3 \times 3$ generations), and very
hierarchical couplings $\lambda \propto y_f^2, y_f^3$ or $y_f^4$,
where $y_f$ is a Yukawa coupling.  
Table \ref{results5} lists some estimates for rare processes
in two cases: quark generation change via CKM, with or
without lepton flavour violation via the lepton mixing matrix.
Processes such as $\tmg$ could be sensitive
to third generation leptoquarks, particularily
for large $\tan \beta$.  If produced
at hadron colliders,  such third generation
leptoquarks would decay to
$t$ or $b$ and $\tau$, or possibly $\mu$. 
However, realistically, leptoquarks of different generations
could be expected to mix, which  could significantly modify the
expectations, due to the steep
hierarchy of couplings. 

\end{enumerate}

In  the tables \ref{results3}, \ref{results4} and
\ref{results5} are listed the $\varepsilon$ factors, for
selected rare processes, which arise for the patterns of
leptoquark couplings considered in  this paper. Although
the expectation for the various patterns differ, various
prospects can be anticipated:
\begin{itemize}
\item  For $\tan \beta = 1$, the expectations  
are well below the  experimental bounds.
So electroweak-scale leptoquarks are possible,  and could
be produced via their gauge interactions at hadron colliders.
The particular collider signatures of each pattern are
briefly discussed at then ends of  sections \ref{add}, \ref{max},\ref{gauge}
and \ref{lqflav}.  Several of the leptoquarks  we consider
decay preferentially to 
 third generation fermions, so leptoquark  searches at
the Tevatron and the LHC for  final states containing
tops and/or taus would be interesting.

\item Lepton flavour 
violating observables, such as  $\tmg, \meg$, and $\mec$,
are sensitive probes of (third generation) leptoquarks, 
because the  leptoquarks can contibute via loops with third
generation couplings, and 
 the experimental bounds on these clean processes are good.

\item Finally, a question for models of $\lambda$ couplings,
is ``which meson decays are most sensitive to leptoquarks?''
Putting aside the D decays, because the experimental bounds
are less restrictive, this amounts to comparing the
predicted branching ratios for $B$  and $K$ decays to
the current bounds. The latter can be several orders 
of magnitude more stringent for $K$s  than for $B$s. 
One can roughly estimate that
 $K$ decays 
may be slightly more sensitive,
when $\lambda \propto y_f$, as can arise
in section \ref{unit}, or if $\lambda \propto \sqrt{y_f y'_f}$.
The patterns discussed in this paper did not give
a square root, but it is 
 expected in the Cheng-Sher ansatz \cite{CS},
and  can arise in various types of models
such as \cite{BG} (expectations with this  ansatz
 also are discussed in \cite{Carpentier}). 
However,  if   $\lambda \propto (y_f y'_f)^n$,
for $ n\geq 1$, as   arises in  most of the patterns
we consider here, then $B$  decays are a better  place to look
for leptoquarks.

\end{itemize}

\section*{Acknowledgements}
We thank Gino Isidori for contributions, and
SD thanks Uli Haisch for a seminar invitation.
This work was partially supported by the EU Contract 
No. MRTN-CT-2006-035482, FLAVIAnet.

\end{document}